\documentclass[fleqn,usenatbib]{mnras}
\usepackage[british]{babel}             
\usepackage{newtxtext}                  
\usepackage{newtxmath}                   
\let\la=\lesssim 
\let\ga=\gtrsim
\usepackage[T1]{fontenc}                
\usepackage{bm}
\usepackage[pdftex]{graphicx}
\usepackage{mathtools}
\usepackage{times}
\usepackage{url}
\usepackage{soul}
\usepackage{float}
\usepackage{caption}
\usepackage[skip=0.5ex]{subcaption}
\usepackage{lipsum}
\usepackage{xcolor}  
\usepackage{comment}
\usepackage{siunitx}
\usepackage{booktabs}
\usepackage{adjustbox}
\usepackage{amsmath}
\usepackage{makecell} 

\usepackage{newtxtext,newtxmath}
\usepackage{xcolor}  

\usepackage[T1]{fontenc}

\DeclareRobustCommand{\VAN}[3]{#2}
\let\VANthebibliography\thebibliography
\def\thebibliography{\DeclareRobustCommand{\VAN}[3]{##3}\VANthebibliography}
\newcommand{\HI} {H\,{\sc i}} 

\usepackage{graphicx}	
\usepackage{amsmath}	
\usepackage{booktabs}   
\usepackage{comment}
\usepackage{subcaption}
\usepackage{hyperref}
\usepackage{tabularx}

\newcommand{\spose}[1]{{\hbox to 0pt{#1\hss}}}
\newcommand{\lta}{\mathrel{\spose{\lower 3pt\hbox{$\mathchar"218$}}
     \raise 2.0pt\hbox{$\mathchar"13C$}}}
\newcommand{\gta}{\mathrel{\spose{\lower 3pt\hbox{$\mathchar"218$}}
     \raise 2.0pt\hbox{$\mathchar"13E$}}}

\newcommand{\kms}{\hbox{km\,s$^{-1}$}}

\title [DINGO/GAMA/WAVES: HIHM]{DINGO / GAMA / WAVES: HI - halo mass relation}

\author[Dev et al.]
{Ajay Dev$^{1}$\thanks{E-mail: ajay.dev@research.uwa.edu.au},
Martin Meyer$^{1}$,
Simon P. Driver$^{1}$,
Jonghwan Rhee$^{1,2}$,
Trystan S. Lambert$^{1}$, \and
Paul Nulsen$^{1,3}$,
Richard Dodson$^{1}$, 
Tobias Westmeier$^{1}$, 
Matthew Whiting$^{4}$, 
Sabine Bellstedt$^{1}$, \and  
Aaron Robotham$^{1}$, 
Jochen Liske$^{5}$,
Elmo Tempel$^{6}$, 
Ivan Baldry$^{7}$,
Jon Loveday$^{8}$, \and
Luke Davies$^{1}$, 
Barbara Catinella$^{1}$, 
Michael J. I. Brown$^{9}$,
Kristine Spekkens $^{10}$, \and
Benne W. Holwerda$^{11}$
\\\\
{}$^1$International Centre for Radio Astronomy Research, The University of Western Australia, M468, 35 Stirling Highway, Perth, WA 6009, Australia.\\
{}$^2$Australia Telescope National Facility, CSIRO Space \& Astronomy, P.O. Box 1130, Bentley, WA 6102, Australia \\
{}$^3$Center for Astrophysics | Harvard \& Smithsonian, 60 Garden Street, Cambridge, MA 02138, USA\\
{}$^4$Australia Telescope National Facility, CSIRO Space \& Astronomy, P.O. Box 76, Epping, NSW 1710, Australia\\
{}$^5$Hamburger Sternwarte, Universität Hamburg, Gojenbergsweg 112, 21029 Hamburg, Germany\\
{}$^6$Tartu Observatory, University of Tartu, Observatooriumi 1, 61602 T\~oravere, Estonia\\
{}$^7$Astrophysics Research Institute, Liverpool John Moores University, IC2, Liverpool Science Park, 146 Brownlow Hill, Liverpool L3 5RF, UK\\
{}$^8$Astronomy Centre, University of Sussex, Falmer, Brighton BN1 9QH, UK\\
{}$^9$School of Physics and Astronomy, Monash University, Clayton, VIC 3800, Australia\\
{}$^{10}$Department of Physics, Engineering Physics and Astronomy, Queen’s University, Kingston, Ontario K7L 3N6, Canada\\
{}$^{11}$Department of Physics and Astronomy, University of Louisville, Natural Science Building 102, Louisville, KY 40292, USA\\
}

\begin{document}
\date{}
\renewcommand{\refname}{REFERENCES}

\pagerange{\pageref{firstpage}--\pageref{lastpage}} \pubyear{2025}

\maketitle

\label{firstpage}

\begin{abstract}
We investigate the relation between neutral atomic hydrogen (\HI) and dark matter halo mass (HIHM) using observations from the Deep Investigation of Neutral Gas Origins (DINGO) pilot survey 100h data, combined with spectroscopic data from the Galaxy and Mass Assembly (GAMA) survey and photometric data from the Wide Area VISTA Extragalactic Survey (WAVES) photometric catalog. We employ a combination of direct detections and spectral stacking to probe the \HI\ content of halos across a wide mass range ($10^{10.5} \lesssim M_\mathrm{h}/M_\odot \lesssim 10^{14.5}$). By incorporating WAVES photometric members on top of the existing GAMA group catalog, we present a novel approach of extending stacking analyses beyond spectroscopic completeness limits, enabling recovery of satellite \HI\ content otherwise missed. We find that the HIHM relation exhibits a double power-law form, with a turnover near $M_\mathrm{h} \sim 10^{11.2} \text{ M}_\odot$. Central galaxies dominate the halo \HI\ budget below $M_\mathrm{h} \sim 6 \times 10^{12} \text{ M}_\odot$, while satellites dominate at higher halo masses. Including photometric members increases the measured \HI\ content in halos above $10^{13} \text{ M}_\odot$ by a factor of 1.5-3, highlighting the importance of gas-rich satellites in the group and cluster regime. Comparison with previous group-stacking studies shows that low-surface brightness galaxies, and intra-group \HI\ structures contribute only a minor fraction to the total \HI\ mass in group and cluster halos, as the summed galaxy \HI\ masses are consistent with the total halo \HI\ content. 

\end{abstract}



\section{Introduction}
The role of cold gas in galaxy evolution is central to our understanding of the baryon cycle of galaxies. Neutral atomic hydrogen (\HI) serves as the primary fuel for star formation \citep[e.g.,][]{Kennicut1998, Saintonge&Catinella2022} and a crucial tracer of gas accretion and environmental processes \citep{Haynes1984, Cortese2021}. In hierarchical structure formation models, galaxies reside in dark matter halos, and the distribution of baryons within and around these halos is shaped by a complex interplay of gas accretion, feedback, mergers, and stripping \citep{White&Rees1978, White&Frenk1991, Somerville2015}. While stellar mass provides an integrated view of past star formation, \HI\ traces a more dynamic phase of galaxy evolution. It is both sensitive to recent accretion and vulnerable to environmental quenching mechanisms \citep[e.g.,][]{Catinella2013, Brown2017}. In low-density environments, some simulations predict that galaxies can accumulate large \HI\ reservoirs through cold mode accretion \citep{Keres2005}, but in dense environments like groups and clusters, processes such as ram-pressure stripping, strangulation, and tidal interactions can rapidly remove or suppress \HI\ \citep{Gunn&Gott1972, Boselli&Gavazzi2006, Cortese2021}.

Quantifying how \HI\ populates dark matter halos is therefore essential to constraining theories of galaxy formation and feedback. The \HI\ –halo mass (HIHM) relation offers a powerful empirical tool in this regard. Analogous to the stellar-to-halo mass relation (SHMR), the HIHM relation connects the cold gas content to the gravitational potential well of the halo. It encapsulates how efficiently halos of different masses can retain or acquire \HI\ and how this efficiency evolves with scale and environment.

The expected form of the HIHM relation is non-linear \citep{Padmanabhan2017}. At low halo masses ($\sim10^{10}$--$10^{11}~M_\odot$), \HI\ is expected to trace the shallow potential wells of gas-rich dwarfs, with high \HI\ -to-halo mass ratios. At intermediate masses ($\sim10^{12}~M_\odot$), corresponding to $L^*$ galaxies and groups, this efficiency may peak. At higher masses ($\gtrsim10^{13}~M_\odot$), corresponding to rich groups and clusters, the \HI\ content is thought to be suppressed due to environmental effects and the transition from cold to hot gas halos \citep{Papastergis2013}.

Observationally, inferring the HIHM relation is challenging. \HI\ detections are limited by sensitivity and typically biased towards gas-rich galaxies. Group and cluster environments, where \HI\ is rarer and more diffuse, are particularly difficult to study with direct detections alone. Recent years have seen the emergence of two key methods to overcome these limitations: (1) stacking of \HI\ spectra to statistically detect \HI\ in populations for which direct detections are not available, and (2) assigning halo masses to galaxies via group catalogs or abundance matching (AM).

\HI\ spectral stacking has been used extensively to quantify the average \HI\ content of galaxy populations \citep[e.g.,][]{Zwaan2000, Chengalur2001, Lah2007, Delhaize2013, Gereb2015, Rhee2013, Rhee2016, Rhee2018,  Bera2019, Chowdhury2020, Chowdhury2022, Sinigaglia2022, Rhee2023, Bianchetti2025}. Several studies have used stacking in conjunction with optical group catalogs (e.g., \citealt{Brown2017, Guo2020, Rhee2023, Hutchens2023, Dev2023}) to explore the HIHM relation. \citet{Guo2020} used a combination of Arecibo Legacy Fast ALFA survey (ALFALFA \citealt{Giovanelli2005}) and Sloan Digital Sky Survey (SDSS, \citealt{York2000}) group catalogs \citep{Lim2017} to perform group spectral stacking to estimate the total \HI\ content in SDSS galaxy groups. A similar group spectral stacking technique was used by \citet{Dev2023}, combining ALFALFA with Galaxy and Mass Assembly Survey (GAMA, \citealt{Driver2022dr4}) group catalog \citep{Robotham2011}, and obtained consistent results with \citet{Guo2020}. \citet{Rhee2023} performed galaxy spectral stacking using Deep Investigation of Neutral Gas Origins (DINGO, \citealt{Meyer2009b}) early science data combined with GAMA  to study the galaxy \HI\ contributions on the HIHM relation. \citet{Hutchens2023} studied the HIHM for sample of very nearby systems using direct detections and estimated the scatter in the relation. 

However, these works are typically restricted to spectroscopic samples and may miss contributions from fainter galaxies within the groups. \citet{Chauhan2021} used \textsc{Shark} \citep{Lagos2018} semi-analytic model to study the HIHM relation by mocking different magnitude limited spectroscopic galaxy surveys. They predict that going from a complete spectroscopic survey for $r<19.8$ to one with a Z-band completeness for $Z<21.1$, can double the total \HI\ mass in the high halo mass bins for a $z<0.1$ selected sample. This highlights the importance of using techniques that account for these fainter systems for obtaining a more accurate estimate of the HIHM relation.

In this work, we study the HIHM relation by combining \HI\ data from the DINGO pilot survey with spectroscopic and photometric galaxy catalogs from GAMA and the Wide Area Vista Extragalactic Survey (WAVES, \citealt{Driver2019}) respectively. All the previous studies of the HIHM relation, except \citet{Rhee2023}, have been using single-dish measurements. Here, we will use the interferometric data from ASKAP (Australian SKA Pathfinder Telescope; \citealt{Johnston2007, Johnston2008, Hotan2021}) which has much higher resolution (DINGO survey has restoring beam size of $30^{\prime\prime}$) than previous generation of radio telescopes (ALFALFA survey on Arecibo had a spatial resolution of $3.5^{\prime}$). We employ both traditional spectroscopic stacking and a novel stacking approach. \textcolor{black}{In the latter, we extend the sample beyond spectroscopically confirmed galaxies by incorporating additional, fainter satellite galaxies drawn from the WAVES photometric catalogue. These photometric satellites are associated with groups based on their proximity to spectroscopically confirmed GAMA centrals. This approach aims to provide a more complete recovery of the \HI\ content in haloes by accounting for contributions from satellite populations that are absent in purely spectroscopic samples.}

Our goals are threefold: (1) to map the HIHM relation across a wide halo mass range using both detections and stacking;  (2) compare our results with past studies to constrain the slope and shape of the HIHM relation; and (3) to assess the impact of including photometric group members on the measured \HI\ content of halos.

The paper is structured as follows: Section 2 describes the datasets. Section 3 outlines our halo mass estimation,  spectral extraction, stacking procedures, and error analysis. Section 4 presents the results from both direct detections and stacks and discusses the implications. We summarize our conclusions in Section 5. Throughout, we assume a Hubble constant of $H_0=70 \text{ km s}^{-1}\text{Mpc}^{-1}$, a $\Lambda$CDM cosmology with
$\Omega_\text{M}=0.3$, $\Omega_\Lambda=0.7$, and Chabrier initial mass function (IMF) for stellar masses \citep{Chabrier2003}.

\section{Data}
\begin{figure}
    \centering
    \includegraphics[width=\linewidth]{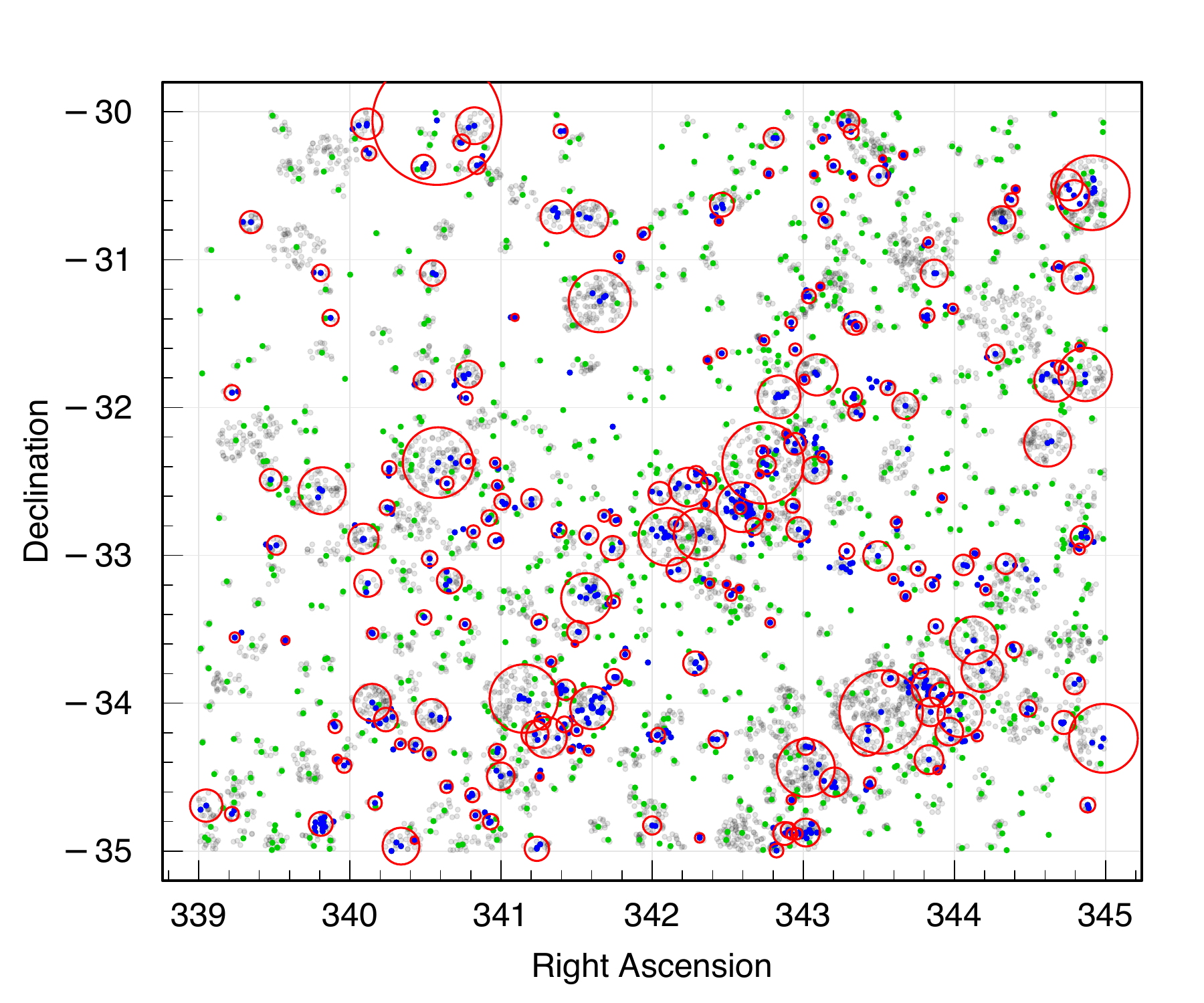}
    \caption{DINGO tile0 footprint and its overlap with the GAMA G23 region. We show the location of GAMA spectroscopic galaxies and groups with $z \le 0.08$. The GAMA galaxies which are part of a group ($N_\text{FoF}>1$) based on G3C are shown in blue while the isolated galaxies are shown in green. The red circles show the group radii corresponding to $R_{200}$ estimated from $M_{200}-R_{200}$ relation. We also show the galaxies from the WAVES photometric catalog with photometric redshift less than 0.2 in grey. 
    }
    \label{fig:footprint}
\end{figure}

\begin{figure}
    \centering
    \includegraphics[width=\linewidth]{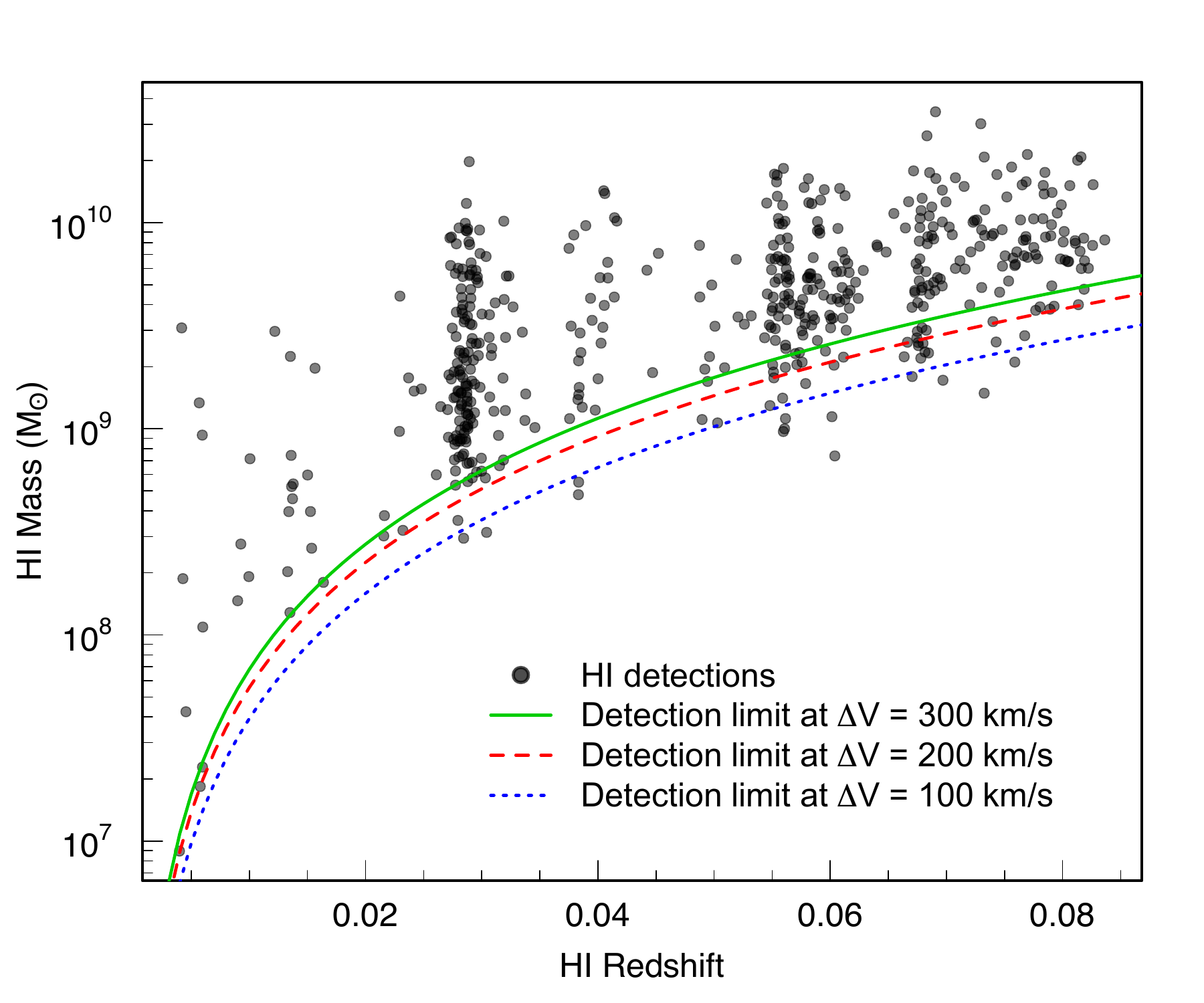}
    \caption{\HI\ mass vs redshift for the DINGO detections. The blue, red and green dashed lines show the DINGO sensitivity curve of a detection with SNR of 7 and velocity widths of 100, 200 and 300 $\text{km s}^{-1}$ respectively.}
    \label{fig:redshift}
\end{figure}

To study total \HI\ content in halos we use a combination of \HI\ data from ASKAP along with an optically-selected sample of galaxies from GAMA/WAVES and galaxy groups from GAMA.

\subsection{DINGO}
The DINGO survey is one of ASKAP’s key survey science programs, designed to trace the evolution of \HI\ in galaxies from the local Universe out to $z \sim 0.4$ \citep{Meyer2009b, Rhee2023}. The main DINGO survey has a target field covering $\sim 60 \text{ deg}^2$ which is divided into two tiles. Each of the tiles will be observed for 800~h in the low-$z$ component ($z < 0.1$) of the survey. In this work, we use the 100-hour pilot survey data from one of the DINGO tiles (hereafter referred to as "tile~0") covering $339^\circ < \text{ R.A. } < 345^\circ $ \& $-30^\circ < \text{Decl.} < -35^\circ$. This region overlaps with the GAMA 23h region. Fig. \ref{fig:footprint} shows the DINGO tile~0 area used in the work. We show the locations of the GAMA spectroscopic galaxies (in green or blue) and the galaxy groups (red circles) which have redshift $z<0.08$. 

DINGO pilot observations were conducted in two phases between 2019 and 2022, utilising the full 36-antenna ASKAP array and a 288 MHz bandwidth. The data were taken in the ASKAP Band 2, spanning 1151.5 to 1439.5~MHz (15552 channels). The resulting channel resolution is 18.52~kHz, which is equivalent to a velocity resolution of $\sim4$~\kms\ at $z=0$. Slightly different observation strategies were applied in Phase 1 and Phase 2.

The ASKAP Phased Array Feeds (PAFs) receiver generates a 6$\times$6 beam configuration, forming an instantaneous field-of-view (FoV) of 30~deg$^2$. The 60~deg$^2$ G23 field was divided into two 30~deg$^2$ tiles. Each tile was covered by two interleaved 6$\times$6 beam footprints to achieve uniform sensitivity across the region. In pilot Phase 1, the two interleaved footprints were alternated every 15 minutes, with one Scheduling Block (SB) covering an entire tile. However, Phase 2 observations separated the two footprints into individual SBs. The integration time for a Phase 2 SB was typically 8 hours, while some Phase 1 SBs had a shorter integration time (refer to Rhee et al. in prep for more details). The total numbers of SBs taken in Phase 1 and Phase 2 are 7 and 6, respectively, corresponding to a total observing time of 100 hours.

The DINGO pilot data were processed on an SB basis using ASKAP\textsc{soft}, ASKAP's science data processing software \citep{Cornwell2011, Whiting2017, ASKAPSOFT, Wieringa2020}. Due to the significant impact of radio frequency interference (RFI) from Global Navigation Satellite Systems (GNSS) below 1295.5~MHz, we used only the higher half of the observed bandwidth, corresponding to 1295.5 to 1439.5~MHz \textcolor{black}{($z<0.1$ for 21 cm emission)}. The processing pipeline followed a standard \HI\ data reduction procedure, including flagging, bandpass/flux calibration, self-calibration, continuum subtraction, and spectral line imaging. More details on DINGO data processing are described in \citet{Rhee2023}.

For the low-$z$ DINGO data cubes, the maximum baseline was limited to 2~km, resulting in a beam size of 30~arcsec. This beam corresponds to a physical size of $\sim 6$~kpc at $z=0.01$ \textcolor{black}{and $\sim 55$~kpc at $z=0.1$}. After spectral line imaging, the 36 beam data cubes of each SB were mosaicked and further combined with the mosaicked data cubes of other SBs, yielding the full 100~h depth data cube. This multi-epoch data combination approach is computationally efficient and straightforward, making it the default option for ASKAP. 

However, it has a limitation in that it can embed artefacts at a level deeper than the individual SB depth, compared to the traditional approach where all data are combined in the \textit{uv} domain and then imaged together. \textcolor{black}{It also limits cleaning depth to the noise level of a single ASKAP scheduling block}. These effects were explored in Rhee et al. (2026, subm.), which found that 92\% of the true integrated flux is recovered by the image domain data combination method used to generate the data cube for this work. Based on this finding, we  apply a correction factor of 1.08 in the subsequent analysis.

With the deep data cube of the median root mean square (RMS) noise of $\sim 0.67$~mJy~beam$^{-1}$~chan$^{-1}$, we performed the \HI\ source finding process using Source Finding Application \citep[SoFiA,][]{Serra2015, Westmeier2021}, an automated 3D source finding software developed for ASKAP \HI\ surveys. From this process, we tentatively identified 506~\HI\ detections (Rhee et al. in prep.). Fig. \ref{fig:redshift} shows the distribution of \HI\ masses as a function of redshift for these detections. We also show the survey sensitivity curves, using the equation for the integrated signal-to-noise (eq. 156)  in \citet{Meyer2017}, for a point source assuming three different galaxy \HI\ velocity profile widths assuming a SNR of 7 and an RMS noise of 0.67 mJy $\text{beam}^{-1}$. 

\subsection{GAMA}
The GAMA survey is a highly complete spectroscopic survey of galaxies down to $r = 19.8$ mag across three equatorial fields (G09, G12, G15), and two non-equatorial fields (G02 \& G23) \citep{Driver2011, Liske2015, Driver2022dr4}. The GAMA dataset offers high-quality redshifts, stellar masses derived from multi-band photometry using spectral energy distribution (SED) fitting \citep{Robotham2020}, and a robust galaxy group catalog \citep{Robotham2011}.

We use the GAMA group catalog for the G23 region (G3CFoFGroupG23v08; referred to as G3C hereafter) to identify central and satellite galaxies, with group properties including multiplicity, velocity dispersion, and projected radius. The G3C was constructed using the spectroscopic sample of galaxies (G3CGalG23v08) using a FoF algorithm as described in \citet{Robotham2011}. Halo velocity dispersions for the groups in G3C are calculated using the Gapper algorithm \citep{Beers1990}. The velocity dispersions are then used to estimate the halo mass for the systems with an accuracy depending on number of group members, ranging from 0.1 dex ($N_\text{FoF}$>10) to 0.3 dex ($N_\text{FoF}$>4).

For this study, we select GAMA galaxies and groups within the redshift range $0.01 < z < 0.08$, which matches the redshift limits of the DINGO data. We use the GAMA spectroscopic redshift information to extract the \HI\ spectra from the DINGO datacube. The stellar masses for the GAMA galaxies, derived with ProSpect SED-fitting code \citep{Robotham2020} using UV to far-IR photometry, are taken from ProSpectAGNv02 catalog \citep{Thorne2022}. We have 1512 GAMA galaxies as part of our sample. 864 galaxies are not assigned to any GAMA group and hence treated as isolated centrals. Out of the remaining 661 galaxies, 218 are group centrals (BCG) and 430 are satellite galaxies. The isolated galaxies and BCGs are together treated as the centrals. The full sample spans a stellar mass range of $7.2 < \text{log}_{10}(M_{*}/\text{M}_\odot) < 11.5$. \textcolor{black}{Note that throughout this paper, we refer to the most massive galaxy in the group as `BCG', irrespective of the halo mass.}

\subsection{WAVES}
The WAVES survey \citep{Driver2019} is a planned spectroscopic program using the 4-metre Multi-Object Spectroscopic Telescope (4MOST, \citealt{deJong2019}) instrument on the VISTA telescope, scheduled to start observations in 2026. WAVES will be targeting over 2 million galaxies with photometric pre-selection from VISTA Kilo-Degree Infrared Galaxy Survey (VIKING, \citealt{Edge2013}) and Kilo-Degree Survey (KiDS, \citealt{deJong2013}) imaging. WAVES includes two main surveys: Wide and Deep. WAVESwide aims to be a highly complete spectroscopic survey (>90 \%) with a limiting Z-band magnitude $\text{Z} = 21.1$, covering $\sim 1150 \text{ deg}^2$ (distributed over 2 fields) at low redshift ($z \le 0.2$). WAVESdeep is aimed at higher redshift ($z \le 0.8$) and planned to go as deep as $\text{Z} \le 21.25$ over $\sim 66 \text{ deg}^2$, split between G23 ($\sim 50 \text{ deg}^2$) and  four deep-drill fields ($\sim 16 \text{ deg}^2$) of the Legacy Survey of Space and Time (LSST, \citealt{Ivezic2019}) chosen to maximize multi-wavelength data from pre-existing surveys.

Source detection for the WAVES photometric catalog was performed using ProFound \citep{Robotham2018} on a detection image generated by stacking the r, i, Z, and Y bands. To address cases of source fragmentation, a process of segment fixing was applied before running ProFound on individual bands (u, g, r, i, Z, Y, J, H, Ks) from VST and VISTA, enabling the derivation of multiband photometry. To enable a robust photometric redshift selection for WAVESwide, four distinct photometric redshift estimation techniques were applied to the photometry catalog: Scaled Flux Matching \citep{Baldry2021}, TOPz \citep{Tempel2025}, GPz \citep{Duncan2022}, and a binary machine learning approach \citep{Kaur2025}. The TOPz, GPz, and SFM estimates are combined using inverse-variance weighting to produce a single overall photometric redshift value ($z_\text{phot\_invar}$) for each source. The machine learning method does not yield a single photo-z estimate in the same way and is therefore not included in this combined value.

For this study, we use the WAVESwide photometric catalog (Bellstedt et al. in prep.) which includes the G23 field. initially applying a photometric redshift cut of $z_\text{phot\_invar} \leq 0.2$, to select likely low-redshift sources and removing any stars or ambiguous sources. Following this selection, we rely solely on the on-sky position (RA, Dec) from the catalog and do not use the photometric redshifts in the subsequent analysis. Any duplicate entries and stars are removed from the sample. With this selection, we have $\sim 30000$ galaxies in the photometric catalog within the tile0 footprint. Almost a third of this sample has spectroscopic redshifts which we use to clean the sample if the spectroscopic redshifts are outside our selection range of $0.01<z<0.08$. After these selections, we have 22735 galaxies in the photometric catalog which we use along with the GAMA sample for further analysis.

\section{Method}

We determine the \HI\ mass in two complementary ways: first, using the direct \HI\ detections from Rhee et al. (in prep.), and secondly, using spectral stacking on the optically selected sample from GAMA and WAVES. We also use two methods to determine the halo masses of the groups: the velocity dispersion-based approach (used in \citealt{Robotham2011}) and abundance matching. We combine these to construct the HIHM relation.

\subsection{Halo Mass}

\subsubsection{Dynamical mass}

The G3C provides velocity dispersion based halo masses, $M_\text{h}=AR\sigma^2$, where $R$ is the radius containing 67\% of the group members, $\sigma$ is the group velocity dispersion and $A$ is the functional scaling factor, which is based on group multiplicity and redshift (MassAfunc column in G3C). The halo mass measurements are available for groups with at least two members, however the mass measurements are only considered robust for groups with 5 or more members \citet{Robotham2011}. Hence, while using the GAMA halo mass based sample, we restrict our study to groups with $N_\text{FoF}>4$. This does restrict this method to high halo masses ($M_h > 10^{12.5} \text{ M}_\odot$).  

\begin{figure}
    \centering
    \includegraphics[width=\linewidth]{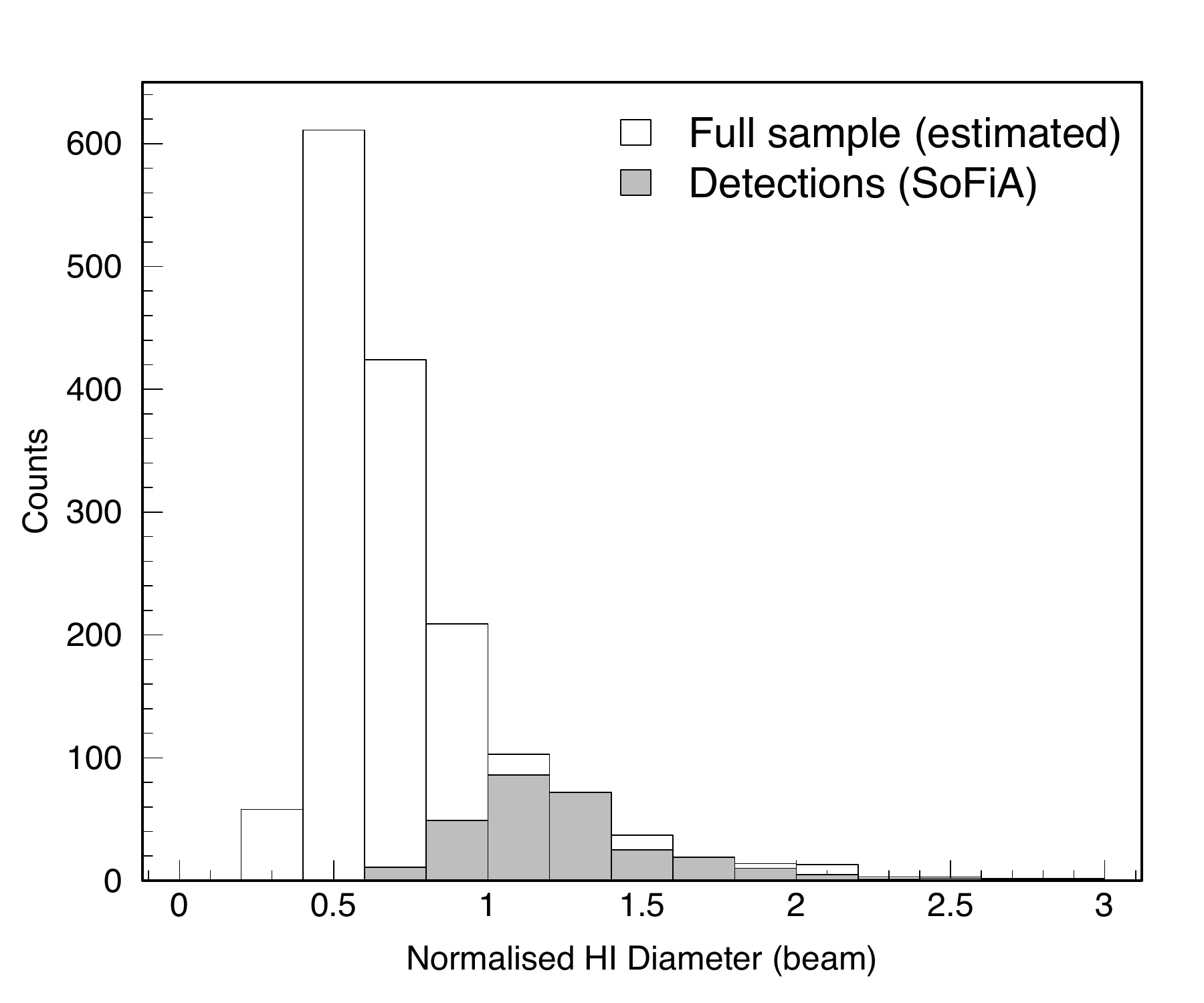}
    \caption{Expected \HI\ size distribution of our GAMA sample normalised to the ASKAP restoring beam size (30"). The shaded region shows the size distribution of the \HI\ detections measured with SoFiA.}
    \label{fig:HIsize}
\end{figure}

\subsubsection{Abundance matching}

AM is a widely used empirical technique for linking galaxies to dark matter halos in the absence of direct mass measurements \citep{Vale&Ostriker2004, Conroy2006, Guo2010, Behroozi2010}. It is based on the assumption of a monotonic relationship between a galaxy property — typically stellar mass or luminosity — and the mass of its host halo, with the assumption of no scatter. By matching the cumulative number densities of galaxies and halos, one can assign halo masses to individual galaxies without the need for dynamical or lensing data. AM has proven successful in reproducing a wide variety of observational statistics, including galaxy clustering, the stellar-to-halo mass relation (SHMR), and satellite fractions \citep{Behroozi2010, Moster2010}. More sophisticated implementations introduce scatter into the $M_\text{star} - M_\text{halo}$ relation or consider subhalo abundance matching, but the fundamental principle remains consistent \citep{Reddick2013}.

For our implementation of abundance matching, we use the galaxy stellar mass function (SMF) for centrals, constructed from the GAMA DR4 galaxy sample following the approach in \citet{Driver2022dr4}. Details of the SMF construction are provided in Appendix~\ref{sec:appendix_gsmf}. The halo mass function (HMF) is adopted from \citet{Driver2022hmf}, which used a combination of galaxy group catalogs and X-ray cluster catalog along with the $\Omega_\text{M}$ constraint from \citet{Planck2018} cosmology. To perform abundance matching, we compute cumulative distribution functions (CDFs) for both the central SMF and HMF by integrating from the highest mass downward. We then match the two CDFs in a one-to-one fashion, assigning each central galaxy a halo mass such that the number density above a given stellar mass equals the number density above the corresponding halo mass. This yields a stellar-to-halo mass relation (SHMR) over the halo mass range $10^{10}$–$10^{15} \mathrm{M}_\odot$.

We assign AM halo masses to all centrals in our GAMA sample based on their stellar masses, which are derived from the Prospector SED-fitting code. The resulting SMHM relation is shown in Fig. \ref{fig:SMHM} (black solid line). Further comparisons of the SMHM relation, including variations in SMFs, HMFs, and comparisons with simulations, are presented in Appendix \ref{sec:appendix_smhm}.

\subsection{HI direct detections}

Rhee et al. (in prep.) found 506 \HI -detected systems with the DINGO 100h pilot survey.  Fig. \ref{fig:redshift} shows the redshift distribution of these detections. 
408 of the detected systems have at least one GAMA optical galaxy within the SoFiA detection mask. 361 of these can be cross-matched uniquely to a GAMA galaxy while 47 detections have multiple optical matches. The remaining 98 systems do not have a GAMA counterpart. However, most of these systems have faint optical counterparts whose magnitudes lie below the GAMA survey magnitude limit. For our stacking analysis, we only include detections within our tile0 footprint that have at least one optical counterpart (338).

\subsection{HI Stacking}
To measure the average \HI\ content of galaxies and their host halos below the direct detection limits of DINGO, we use the galaxy spectral stacking technique. \textcolor{black}{This approach significantly boosts the signal-to-noise ratio of \HI\ spectra by averaging over large samples, with the noise in the stacked spectrum decreasing approximately as $\sqrt{N}$. As a result, stacking enables measurements well below the single-object detection threshold and allows the average \HI\ properties of galaxy populations to be robustly recovered, even in regimes where the majority of individual spectra are too weak for direct detection.} We do this by extracting \HI\ spectra from the DINGO datacube at the positions of optically selected galaxies, aligning them in rest-frame velocity space, and co-adding the spectra within halo mass bins. The stacking procedure weights each frequency channel by the inverse square of its RMS noise ($w = \sigma_\text{RMS}^2$) when combining the \HI\ spectra \citep[e.g.,][]{Chengalur2001, Lah2007, Bera2019, Delhaize2013, Sinigaglia2022, Rhee2023}. 

\subsubsection{Spectral Extraction}
For each optical source in our sample, we extract a spectrum within a 5$\times$5 pixel aperture (pixel size - 6") from the DINGO data cube, centered on the galaxy location specified in the optical catalog. This aperture is chosen as it roughly covers one beam (30"$\times$30"). The spatially-integrated flux density for each spectral channel is extracted following Eq. 8 in \citet{Shostak&Allen1980}
\begin{equation}
    \text{S}_{v} = \frac{\Sigma_x\Sigma_y\text{S}_{v}(\text{x,y)}}{\Sigma_x\Sigma_y\text{B}(\text{x,y)}},
\end{equation}

where x, y are the pixel coordinates of a galaxy and B(x,y) is the normalised beam response of the ASKAP PSF at the position (x,y), which can be approximated as 2D elliptical gaussian. \citet{Rhee2023} estimated the expected \HI\ sizes of galaxies in the DINGO sample from the G23 region using a scaling relation between the \HI\ size and optical B-band luminosity,
\begin{equation}
    \text{log}_{10}D_\text{\HI} = -0.123M_\text{B} - 0.927.
\end{equation}
We use this equation to estimate the \HI\ sizes of our sample using the B-band magnitude obtained from the g-r colors, following the transformation equation in \citet{Jester2005}. Fig. \ref{fig:HIsize} shows the estimated \HI\ sizes for our GAMA sample normalised with the beam size. Majority of our galaxies have \HI\ sizes smaller than one beam, hence the 5$\times$5 aperture is sufficient to capture the flux from our sources. For the case of detections, we do not perform a spectral extraction, but instead use the spectra provided by SoFiA.

\begin{figure*}
    \centering
    \includegraphics[width=0.9\linewidth]{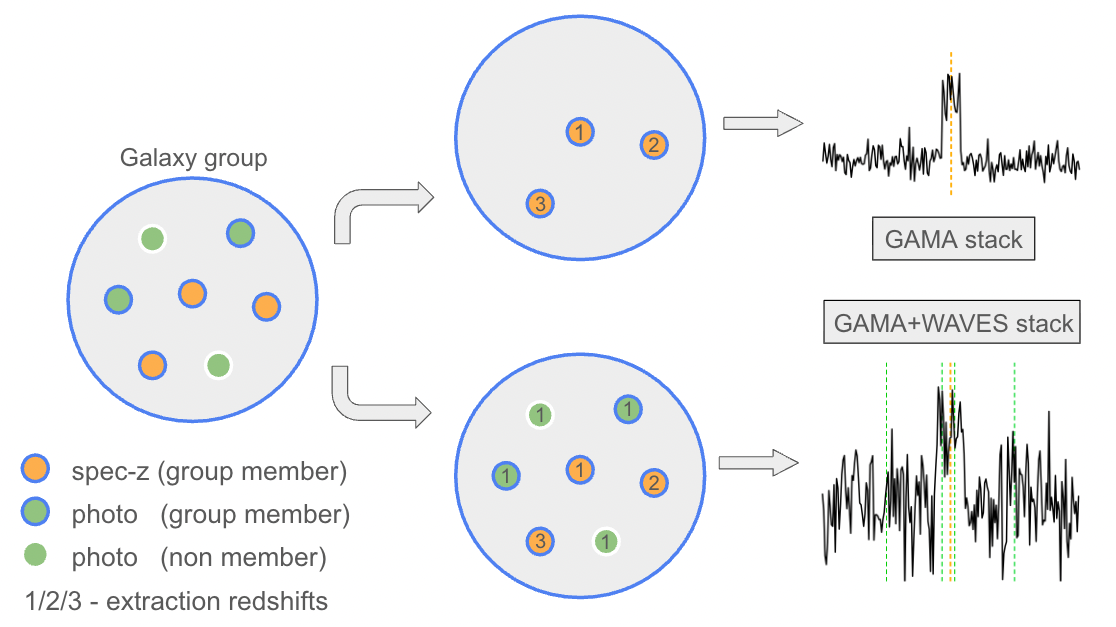}
    \caption{Schematic of the two types of stacking done in this work. First is the "GAMA stacking", where we extract the DINGO spectra of galaxies at the RA, Dec and spectroscopic redshift of each of the group members. Second is the "GAMA+WAVES stacking", where we use the WAVES galaxy photometric catalog on top of the GAMA spectroscopic catalog. Galaxy group members from the GAMA+WAVES sample are taken to be any galaxy within the projected $R_{200}$ of a GAMA group central. Only a tiny fraction of the group satellite members from the GAMA+WAVES catalog have spectroscopic redshifts. Hence, we extract the DINGO spectra using the same approach as GAMA stacking for group members that have a spectroscopic redshift. While for group members that do not have a spectroscopic redshift, we extract the DINGO spectra at the RA and Dec of the GAMA+WAVES galaxy location but at the redshift of the group central (which does have a spectroscopic redshift). This approach will help to capture the signal from any photometric galaxy that happens to be a member of the group, while just adding noise if the galaxy is in projection. The orange shaded circles indicate galaxies which have a spectroscopic redshift, while the green shaded circles indicate galaxies that do not have spectroscopic redshift. The presence of a blue circle boundary indicates that the galaxies are part of the galaxy group while the rest are galaxies in projection. The numbers indicate the redshift used to extract the spectra. The orange dotted lines on the stacked spectra indicate the location of spectroscopic galaxies which have been shifted to rest frame and aligned correctly. The green dashed lines indicate the location of the galaxies which do not have spectroscopic redshift and instead used the redshift of the central galaxy to align with the spectroscopic sample. Hence a small deviation between orange and green line indicates redshift difference between a central and satellite while a large offset indicates a galaxy in projection.}
    \label{fig:schematic}
\end{figure*}

The observed spectra are shifted to the rest frame of the galaxy using the optical spectroscopic redshift, ensuring that the flux is conserved. The flux density is converted to \HI\ mass per unit frequency using the equation in \citet{Delhaize2013},
\begin{equation}
\frac{M_{\mathrm{HI},\nu}}{\text{M}_\odot\,\mathrm{MHz}^{-1}} = 4.98 \times 10^7\, \left(\frac{S_\nu\ }{\mathrm{Jy}}\right) \left(\frac{D_L}{\mathrm{Mpc}}\right)^2,
\end{equation}
where $S_\nu\ $ is the \HI\ flux density in the rest frame and $D_L$ is the luminosity distance to the galaxy estimated from the optical redshift \citep{BryanNorman1998}. All the rest-frame \HI\ spectra are then re-binned to a common frequency grid for stacking. We perform stacking with multiple configurations that accommodate variations in halo mass estimation and the addition of galaxies with only positional information. These aspects are detailed in the following subsections. In Fig. \ref{fig:schematic}, we schematically show the two different stacking approaches followed in this work - first using the spectroscopic sample only and the second by combining the spectroscopic sample with an additional sample of galaxies having only photometric information. 

We also account for any confusion in our sample. If two or more GAMA group galaxies are within a beam, we count them only once in our stack. Less than 2\% of the sample have this correction. For GAMA galaxies, which have a DINGO detection, we use the SoFiA spectra instead of the spectra extracted using the $5\times5$ pixel aperture. This ensures that we are not missing any flux from our detected sources which can cover multiple beams (see the shaded region in Fig. \ref{fig:HIsize}). Some of the DINGO detections have multiple optical GAMA counterparts. In such cases, we count the detection only once in the group halo to avoid overcounting. The SoFiA masks would need further cleaning to separate the emission due to individual galaxies, however this does not cause an issue for us as we are interested in the total \HI\ in all galaxies in the group. 

\subsubsection{GAMA Spectroscopic Sample with Velocity Dispersion-Based Halo Masses}

Our first stacking configuration (Fig. \ref{fig:schematic}, top row) is based on galaxies that have spectroscopic redshifts from GAMA and are within the DINGO footprint. We select galaxies which are assigned to a galaxy group in the G3C. The group halo masses, obtained from G3C, are based on the line-of-sight velocity dispersion of group members, and are more reliable for groups with sufficient richness. We restrict our stacking to groups with $N_\text{FoF} > 4$ to ensure the robustness of the inferred dynamical masses.

Galaxy groups from the G3C are binned in logarithmic halo mass bins of width 0.5 dex. We select the rest-frame \HI\ spectra of all the member galaxies of the groups within each bin. These are co-added to create a stacked spectrum. We perform linear baseline subtraction on the stacked spectrum to remove any residual baseline offset. The linear baseline is calculated by performing a linear fit within a 300 $\text{km s}^{-1}$ region on either side of velocity integration window. The baseline is then interpolated across the integration window and subtracted from the stacked spectrum. The velocity integration window is chosen to account for the expected rotational velocity of galaxies, following the scaling relations presented in \citet{Li2019} and assuming an NFW halo profile. The total \HI\ mass is then obtained by integrating the stacked spectrum over this velocity window. 

The total \HI\ mass derived from the stack is normalized by
the number of distinct groups contributing to the bin, providing an estimate of the average group \HI\ mass for a given halo mass bin.

\begin{figure}
    \centering
    \includegraphics[width=\linewidth]{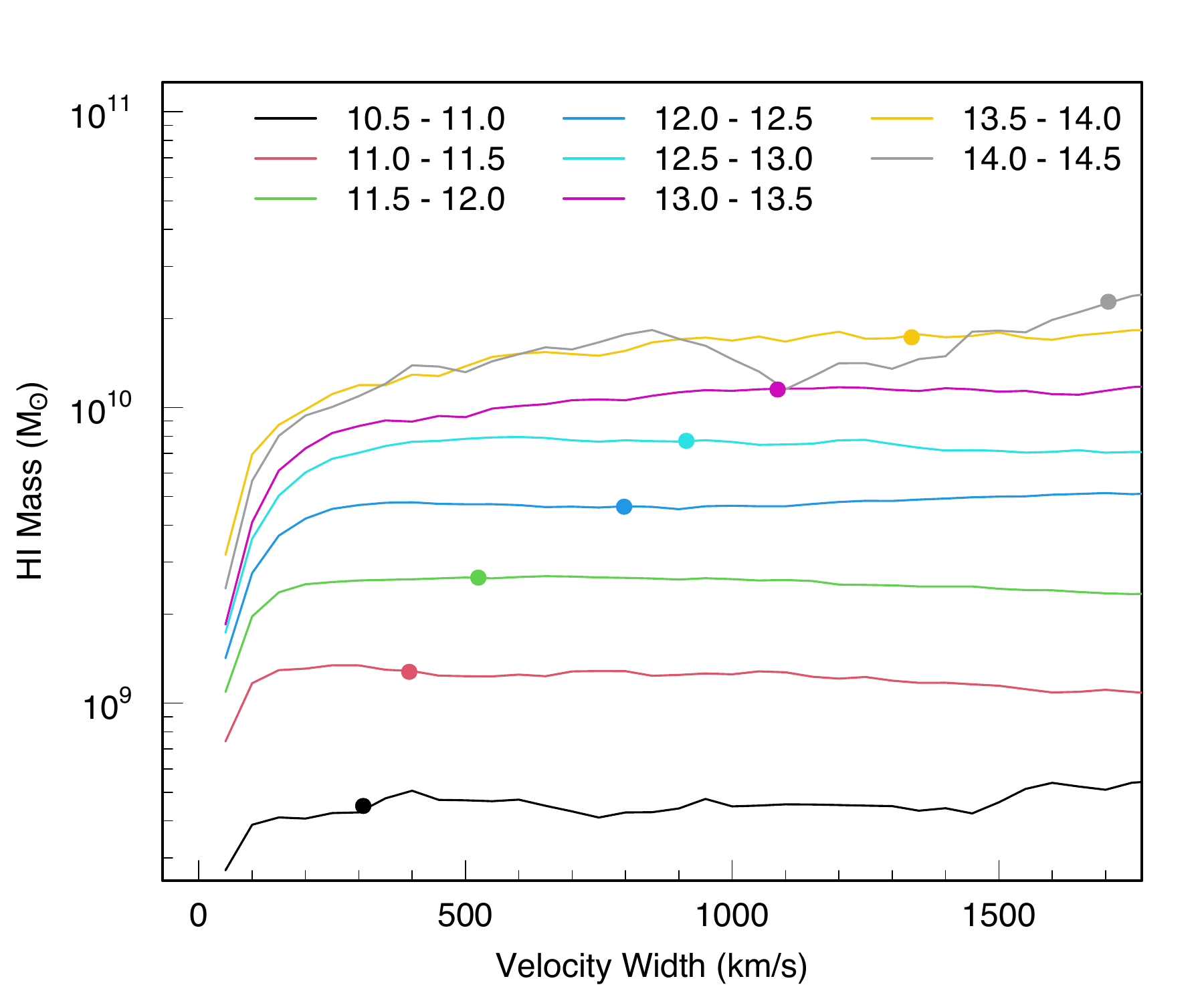}
    \caption{Curve of growth - \HI\ mass as a function of velocity width for the GAMA+WAVES AM-based stacks in different halo mass bins. The dots show the velocity widths used for integration for each of the halo mass bin which is proportional to the halo velocity dispersion.}
    \label{fig:CofG}
\end{figure}

\subsubsection{GAMA Spectroscopic Sample with Abundance Matching-Based Halo Masses}

In the second configuration, we use the same GAMA spectroscopic sample but assign halo masses to central galaxies through stellar mass-based AM. This approach involves a monotonic mapping between the cumulative SMF and the HMF, as detailed in Section~3.1, and is independent of group membership or velocity dispersion measurements.

Using this method allows us to include both isolated centrals and galaxy groups with low-richness that would otherwise be excluded from velocity dispersion-based stacking. The GAMA spectroscopic galaxy sample is divided into bins of 0.5 dex based on the AM-assigned halo masses. For each galaxy, we stack the rest-frame-aligned \HI\ spectra within each bin. The integration of the stacked spectrum follows the same methodology as above. 

This configuration enables us to probe the average \HI\ content of halos across a broader halo mass range, from $10^{10.5}$ to $10^{14.5}\,\text{M}_\odot$, and complements the velocity dispersion-based method by including the low-mass and low-richness regimes inaccessible to the former.

\begin{figure*}
    \centering
    \includegraphics[scale=0.4]{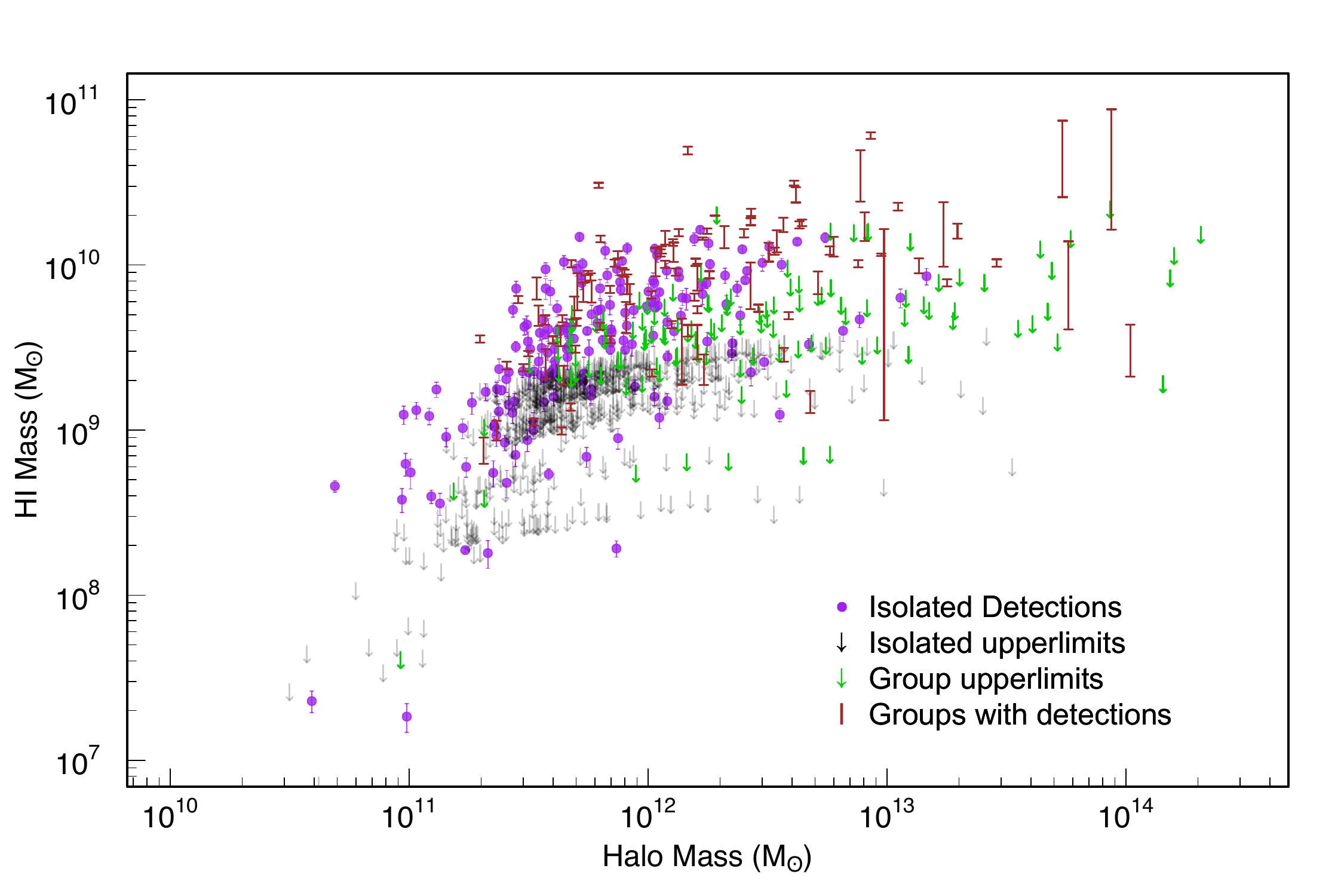}
    \includegraphics[scale=0.4]
    {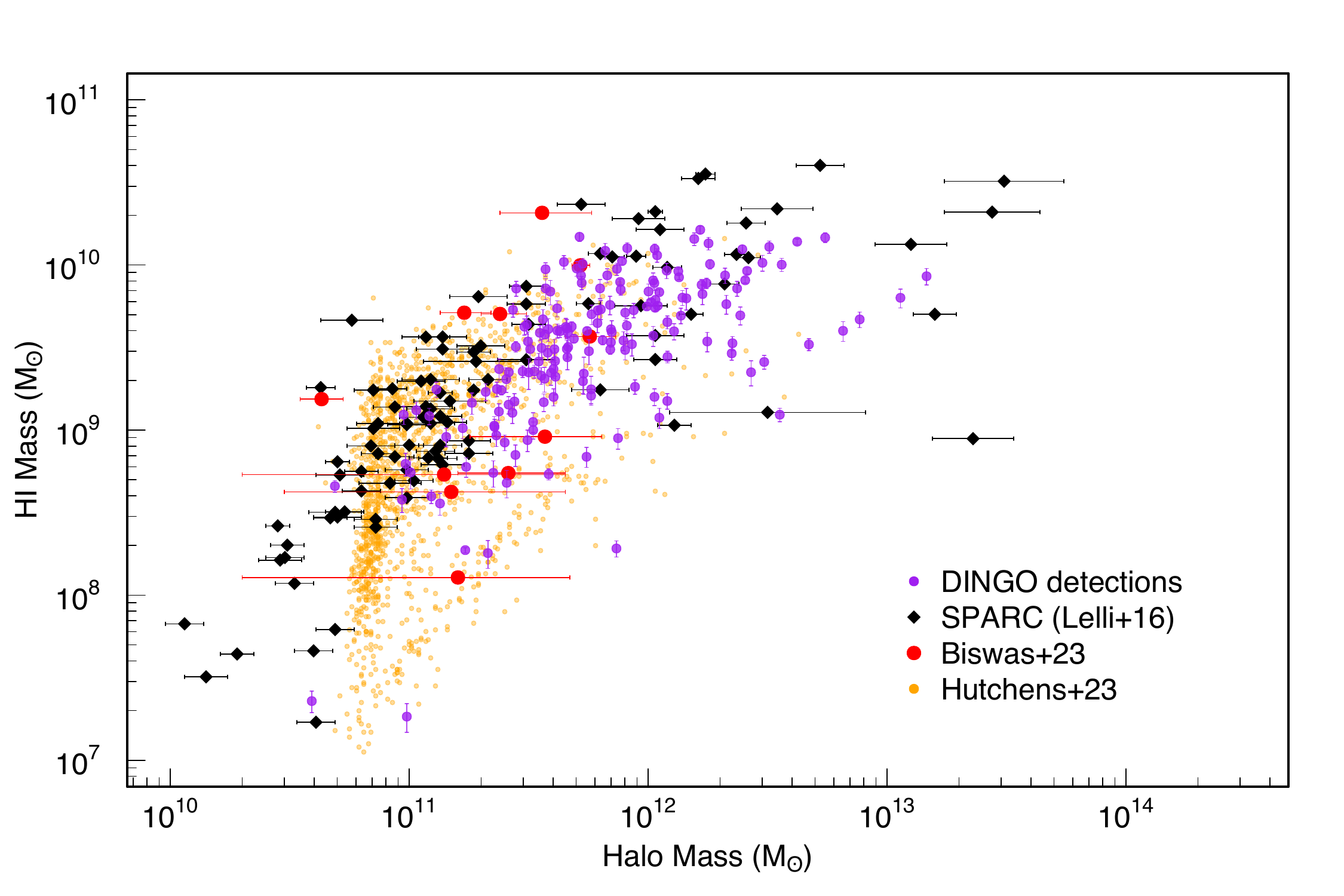}
    \caption{Top: DINGO isolated galaxy detections along with upper limits based on DINGO sensitivity. For isolated GAMA galaxies (not part of any GAMA group), we show the upper limits in black. In green, we show the group upper limits for groups without any detections and we show the groups with at least one detection in brown. The halo masses for these systems are obtained through AM. Bottom: DINGO detections on the HIHM plot along with comparisons with data from SPARC \citep{Lelli2016}, \citet{Biswas2023} and \citet{Hutchens2023}.}
    \label{fig:Detections}
\end{figure*}

\subsubsection{Inclusion of the WAVES Photometric Sample}

\textcolor{black}{Although the GAMA spectroscopic survey achieves high completeness, it is still limited by its magnitude threshold. As a result, galaxy groups may contain fainter satellite galaxies that fall below the spectroscopic limit and therefore lack reliable redshift measurements. Stacking analyses based solely on spectroscopically confirmed group members would therefore miss the \HI\ contribution from these faint satellites, potentially leading to an underestimate of the total \HI\ content of the halo. While deep optical imaging data exist for these galaxies, their photometric redshift uncertainties are typically too large to allow them to be directly incorporated into spectral stacking.}

To extend our analysis to galaxies not covered by GAMA spectroscopy, we incorporate additional sources from the WAVES photometric catalog (Bellstedt et al. in prep.). This catalog provides high-precision photometry and positions, enabling us to include galaxies without spectroscopic redshifts while mitigating the impact of redshift uncertainty through group association. A fraction of the WAVES galaxies would have spectroscopic redshifts from previous surveys on the same field such as GAMA, however the majority only have photometric redshifts. We use this to make a photometric redshift cut ($z_\text{phot} < 0.2$) to create a subsample of WAVES galaxies to be used in our stacking. \textcolor{black}{This cut is motivated primarily by the redshift range of the DINGO 100h data used in this work ($z \le 0.08$), and serves to reduce the number of high-redshift projected interlopers that would otherwise increase the noise in the stacked signal. The mean photometric redshift uncertainty for the this sample of WAVES galaxies is $\Delta z/z \sim 0.026$, which is small compared to the width of the redshift cut and therefore does not significantly bias the selection. In future implementations, this selection could be further refined by incorporating the full photometric redshift probability distribution functions (PDFs) to weight galaxies by their likelihood of association within a redshift window, thereby optimising the balance between completeness and contamination. We note that we do not use the photometric redshifts for any other steps including stacking.} 

The next step is to associate the WAVES galaxies to galaxy groups. First we identify centrals — either the group centrals or isolated centrals — from G3C, ensuring that they have spectroscopic redshifts and stellar mass estimates. These centrals are present in the WAVES catalog as well. Satellite galaxies already associated to a central through G3C are directly incorporated. Additional potential satellite galaxies from the WAVES catalog (which do not have spectroscopic redshifts) are then associated with these centrals if they are within a projected separation of $R_{200}$ from the central. The radius for a particular group is computed using an $M_{200}-R_{200}$ scaling relation \citep{BryanNorman1998}. 
We will refer to this final catalog with GAMA centrals, existing GAMA satellites and additional WAVES galaxies as the GAMA+WAVES catalog hereafter. The total number of galaxies in the GAMA+WAVES sample after including the satellites from WAVES to the GAMA centrals is 8806. The number of galaxies and groups in each of our samples described above is  provided in Table. \ref{tab:halo_counts}.

Once the group membership is assigned, we extract the DINGO \HI\ spectra for all the galaxies in our GAMA+WAVES catalog. Wherever a spectroscopic redshift is available, we use it to shift the spectra to the rest frame. For galaxies without a spectroscopic redshift, we instead adopt the spectroscopic redshift of the associated central galaxy. This approximation is motivated by the expectation that true satellite galaxies occupy the same halo and therefore lie close to the group systemic velocity, with offsets set primarily by the halo velocity dispersion. In this configuration, satellites without spectroscopic redshifts will therefore contribute a broadened but still coherent signal around the central redshift, while galaxies that are only projected within $R_{200}$ and are not physically associated with the group will be shifted incorrectly and will contribute primarily as additional noise rather than coherent signal in the stack. \textcolor{black}{From the spectroscopic sample, the typical velocity offsets between centrals and satellites range from $\sim100 \text{ km s}^{-1}$ at $M_\mathrm{h}\sim10^{11}\,\mathrm{M_\odot}$ to more than $1000 \text{ km s}^{-1}$ at $M_\mathrm{h}\sim10^{14}\,\mathrm{M_\odot}$, broadly tracing the increase in group velocity dispersion with halo mass. To account for this effect, the velocity integration window includes the expected group velocity dispersion, a component corresponding to the rotational velocity of the outermost \HI\ gas in a satellite estimated using the scaling relations of \citet{Li2019}, and an additional conservative buffer of $150 \text{ km s}^{-1}$. This integration window is chosen to encompass the signal from almost all true photometric group members if their velocity distribution is similar to that of the spectroscopic sample.} The rest of the stacking process follows the same methodology as described in Section 3.2.2 and 3.2.3. This stacking configuration is schematically displayed in Fig. \ref{fig:schematic} (bottom row).

\begin{figure*}
    \centering
    \includegraphics[width=\linewidth]{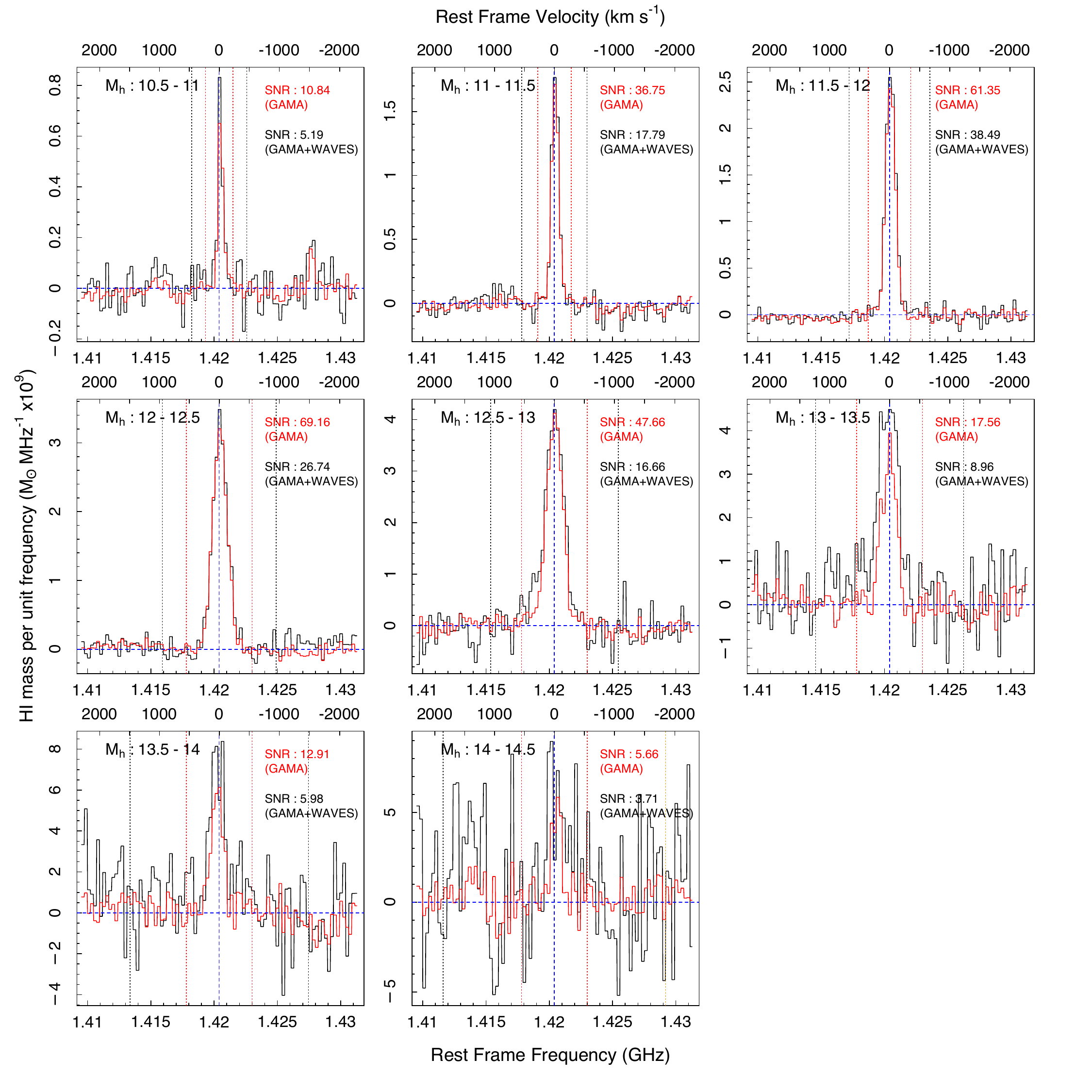}
    \caption{Stacked spectra using AM GAMA (red) and AM GAMA+WAVES (black) catalog for 9 halo mass bins.  The red and black dotted lines show the frequency width used for integrating the total \HI\ mass for the AM GAMA and AM GAMA+WAVES catalog respectively. The vertical blue dotted  lines denotes the rest frame frequency. In each panel, we also show the halo mass bin, and the SNR for the GAMA and GAMA+WAVES stacks.}
    \label{fig:spectra}
\end{figure*}

In Fig. \ref{fig:CofG}, we show the curve of growth for our stacks based on the AM GAMA+WAVES catalog. Each curve shows the integrated \HI\ mass for a given halo mass bin as a function of the velocity width of integration. The curve is expected to rise and then flatten (or have a small slope) once most of the \HI\ contribution from a particular group is captured, with  variations thereafter arising from noise or galaxies in projection which are not part of the group. The colored dots show the final velocity width of integration used for different halo mass bins, and are in the flat part of the curve of growth.

For the WAVES photometric galaxies added in this approach, if they are true group members, their \HI\ emission will contribute to the stacked signal. \textcolor{black}{However, because the spectra are extracted at the redshift of the central galaxy, the emission from these satellites will generally be offset by their line-of-sight peculiar velocities relative to the group centre. This introduces an additional broadening to the stacked profile that reflects the velocity distribution of satellites within the halo. The magnitude of this effect is expected to scale with halo mass through the group velocity dispersion, although the effective broadening will be smaller than the full dispersion because spectroscopic members are already aligned in redshift space.} On the other hand, if the added WAVES source is a projected interloper, its emission will not align in velocity space and will therefore increase the noise in the stacked spectrum rather than the signal. As a result, this method allows us to recover additional \HI\ signal from satellites that might otherwise be missed due to their absence from the spectroscopic sample, albeit at the potential cost of degraded signal-to-noise in the stacked spectrum. This novel approach therefore provides a more complete census of the total \HI\ content of galaxy groups.

\subsection{Error Analysis}
We estimate the uncertainties in our stacked \HI\ mass measurements using a comprehensive Monte Carlo (MC) framework that accounts for both observational and model-based sources of error. First, for the velocity dispersion based halo masses, we incorporate a conservative uncertainty of 0.3 dex on the halo mass (see Fig. 3 in \citealt{Driver2022hmf}). This is implemented by perturbing the velocity dispersion derived halo masses assuming a log-normal scatter. For halo masses derived via abundance matching, we propagate uncertainties by varying both the SMF and the HMF within their reported uncertainties \citep{Driver2022dr4, Driver2022hmf}. At each iteration, we regenerate the stellar-to-halo mass relation and reassign halo masses to central galaxies accordingly. In addition, we perturb the stellar masses of the central galaxies within their respective measurement errors, and recalculate the corresponding abundance-matched halo masses. This captures the systematic uncertainty associated with the construction of the SHMR itself and the scatter in galaxy stellar mass estimates.

To account for sample variance, we perform a MC resampling of the galaxies in each halo mass bin. In each MC iteration, we randomly select 90\% of the galaxies in the bin and apply the full stacking pipeline. This step captures the impact of stochastic sampling and intrinsic scatter in the \HI\ properties of galaxies within each bin.

We perform a total of 100 such MC realizations. For each halo mass bin, we compute the mean and standard deviation of both the stacked \HI\ mass and the corresponding halo mass. These are reported as the final measurements and statistical uncertainties in our HIHM relation. Finally, for each MC iteration, we measure the RMS noise from line-free regions of the stacked spectrum and propagate it into the uncertainty on the integrated \HI\ flux. The noise contribution is combined in quadrature with the MC-derived scatter to obtain the total error on the \HI\ mass.

This approach ensures that our reported uncertainties reflect a realistic combination of measurement errors, sampling variance, and systematic uncertainties in halo mass estimation. However, we also note that there is an inherent systematic arising from the choice of HMF used to construct the AM relation. This effect is discussed in Appendix \ref{sec:appendix_smhm}, where we used various HMF to construct the AM stellar to halo mass relation.

\section{Results \& Discussion}

\begin{figure}
    \centering
    \includegraphics[width=\linewidth]{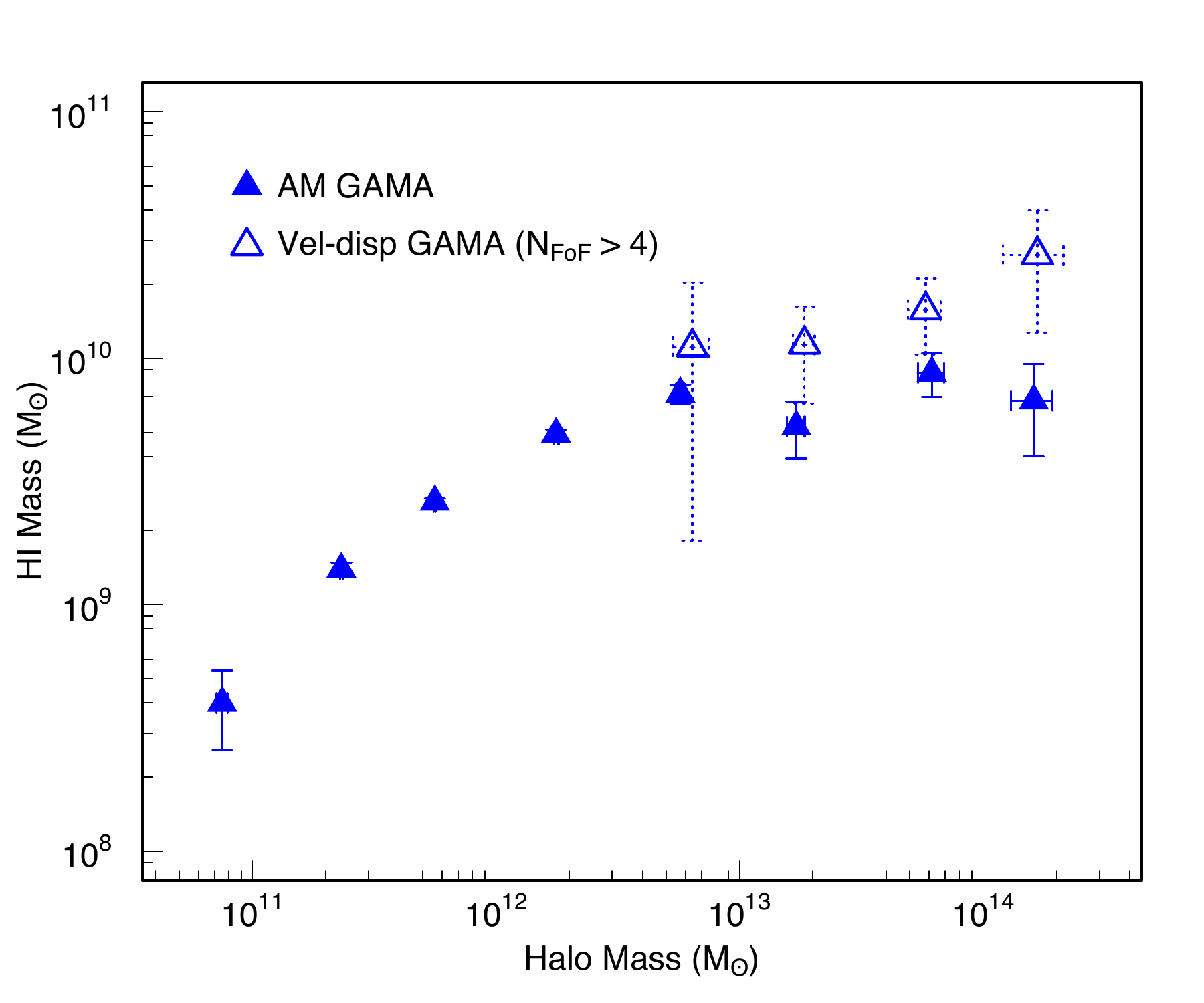}
    \caption{GAMA spectroscopic sample based stacking results on the HIHM plane. Open triangles denote halo masses determined from the velocity dispersion for GAMA groups with multiplicities greater than 4, while filled triangles denote halo masses determined by abundance matching.}
    \label{fig:HIHM_GAMA}
\end{figure}

\begin{figure*}
    \centering
    \includegraphics[width=0.49\linewidth]{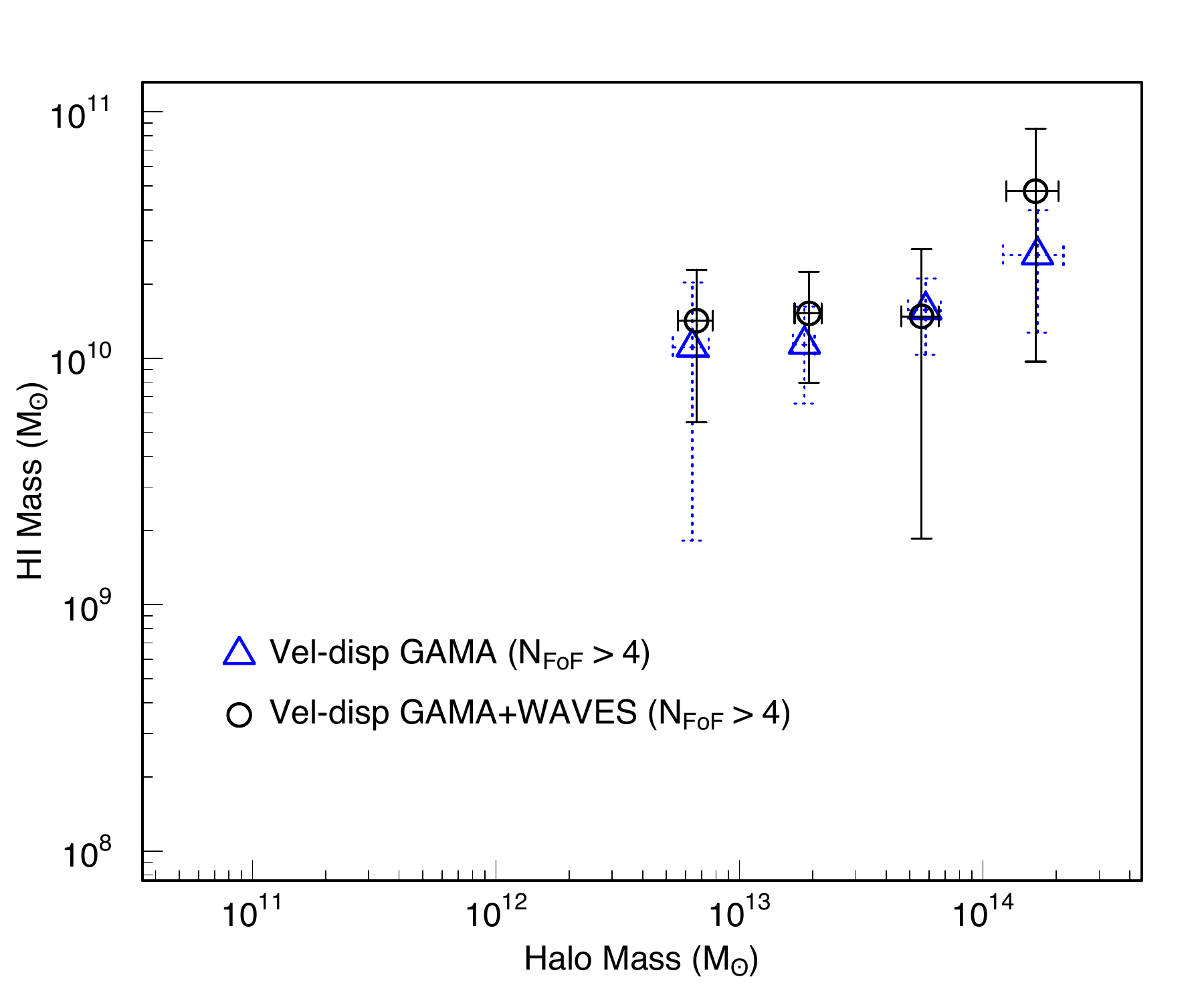}
    \includegraphics[width=0.49\linewidth]{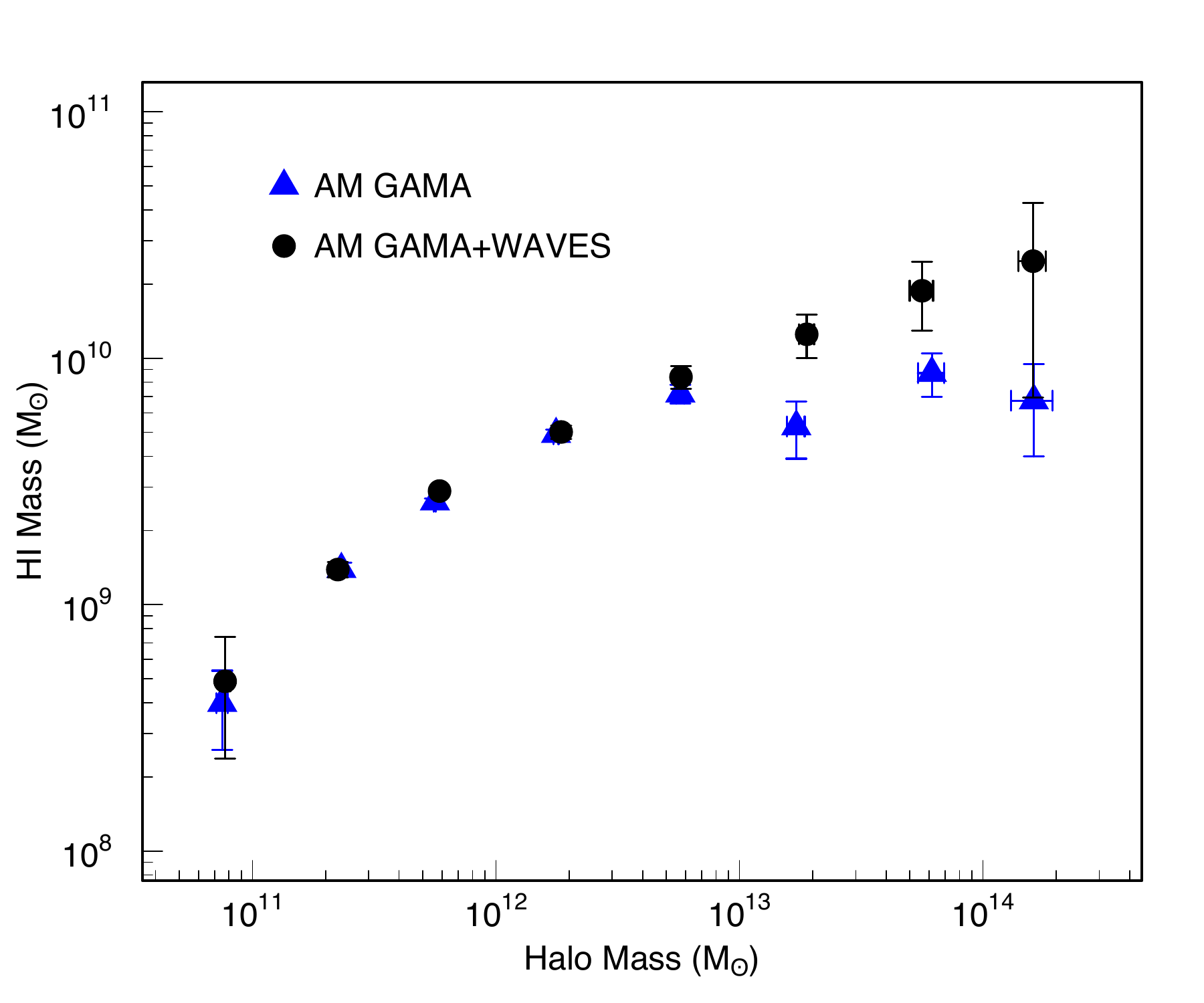}
    \caption{Left: HIHM using velocity dispersion based halo masses with GAMA and GAMA+WAVES for systems with at least 4 spectroscopically detected group members. Right : HIHM using AM based halo masses with GAMA and GAMA+WAVES sample. }
    \label{fig:HIHM_waves}
\end{figure*}

\begin{figure*}
    \centering
    \includegraphics[width=\linewidth]{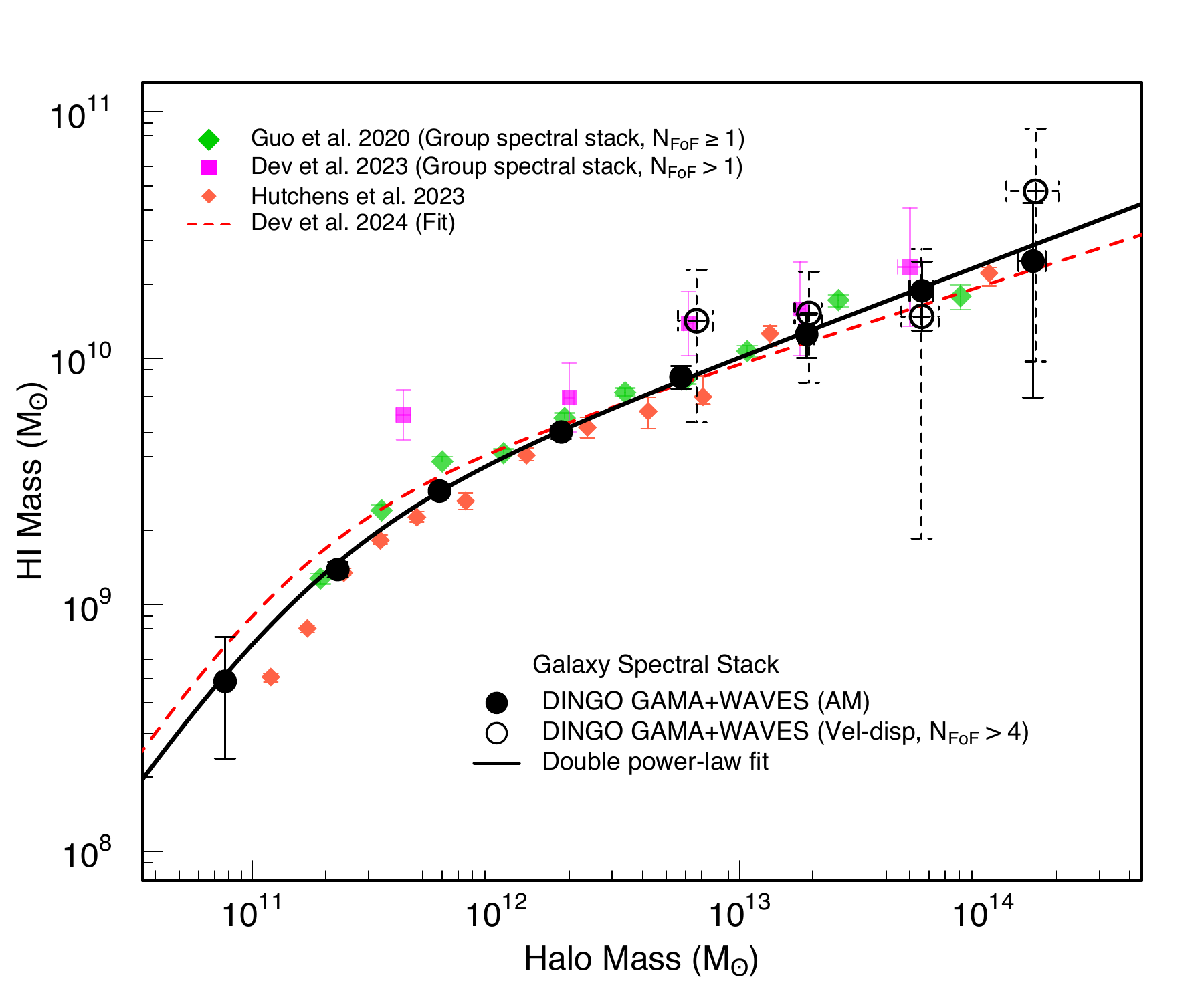}
    \caption{GAMA+WAVES photometric catalog based stacking results with AM halo masses (black) compared with group spectral-stacking studies in the literature (\citealt{Guo2020, Dev2023}). The double power-law fit for our AM based datasets is shown in black. We also show a double power-law fit (red) of the HIHM relation from \citet{Dev2024} based on a compilation of detection and stacking based results. }
    \label{fig:HIHM_comparison}
\end{figure*}

\begin{figure}
    \centering
    \includegraphics[width=\linewidth]{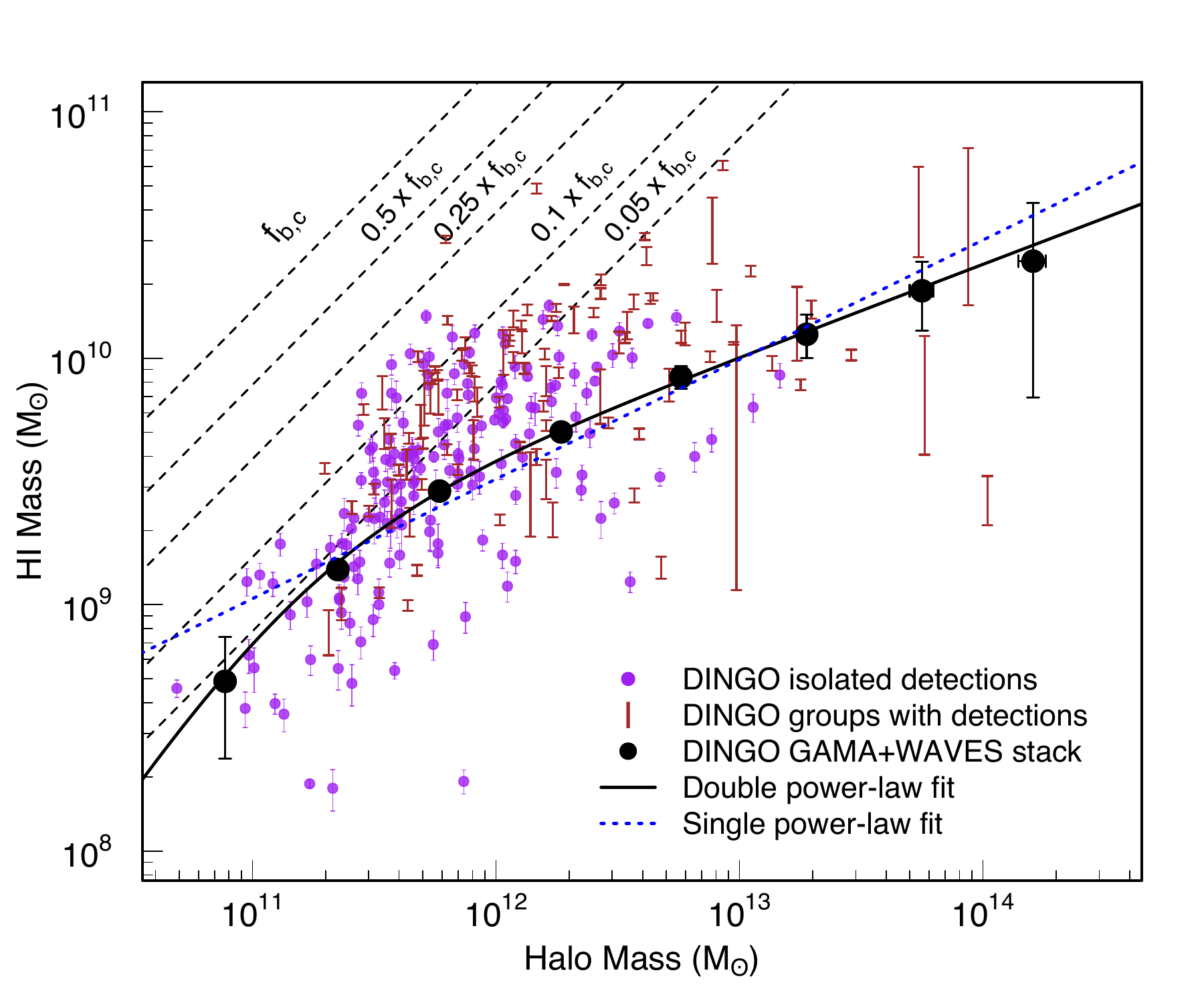}
    \caption{DINGO detections, and stacks based on AM based halo mass using the GAMA+WAVES sample. The black solid line shows the best-fit double power law model for the stacks and the blue dotted line shows the best-fit single power law. The dashed lines show the cosmological baryon fraction and fractions of it of 50\%, 25\%, 10\% and 5\%.}
    \label{fig:HI_fb}
\end{figure}

\subsection{Direct detections}
 
Fig. \ref{fig:Detections} (top) shows the DINGO isolated detections along with upper limits on the isolated non-detections estimated based on their redshift and DINGO sensitivity curve assuming a SNR of 7 and velocity width estimated based on the \HI\ Tully-Fisher relation in \citet{Catinella2023}. For GAMA groups that do not have any detections, we sum the \HI\ mass upper limits on each group member and get an upper limit estimate on the total group \HI\ mass. These are shown in green. For groups which have at least one detection, the \HI\ mass of the detections can be used as a lower limit for the group \HI\ content. These are combined with the upper limits from non-detections to obtain the group \HI\ mass measurements shown in brown.  

The detections and the upper limits can be used to constrain the scatter in the HIHM relation. The upper end of the detections with total group upper limits can roughly be used to trace the upper envelope of the HIHM relation.  \citet{Hutchens2023} studied the HIHM using \HI\ detections and upper limits from RESOLVE \citep{Kannappan&Wei2008} and ECO \citep{Moffett2015} surveys. Our mean relation based on the detections agrees with their mean relation. They also obtained a similar or slightly higher spread in the relation, especially for the lower halo masses (see their Fig. 17). 

Fig. \ref{fig:Detections} (bottom) shows the isolated detections in the HIHM plane, where the halo masses are estimated using abundance matching. Plotted alongside are the Spitzer Photometry and Accurate Rotation Curves (SPARC) sample \citep{Lelli2016, Li2019} Giant Metrewave Radio Telescope (GMRT) based measurements from \citet{Biswas2023}. For these measurements, halo masses were estimated using rotation curve modelling. Also included are measurements from \citet{Hutchens2023} where halo masses are based on abundance matching. DINGO detections sample the same distribution as these previous measurements in the HIHM plane.

\subsection{Stacking results}

\begin{figure*}
    \centering
    \includegraphics[width=0.49\linewidth]{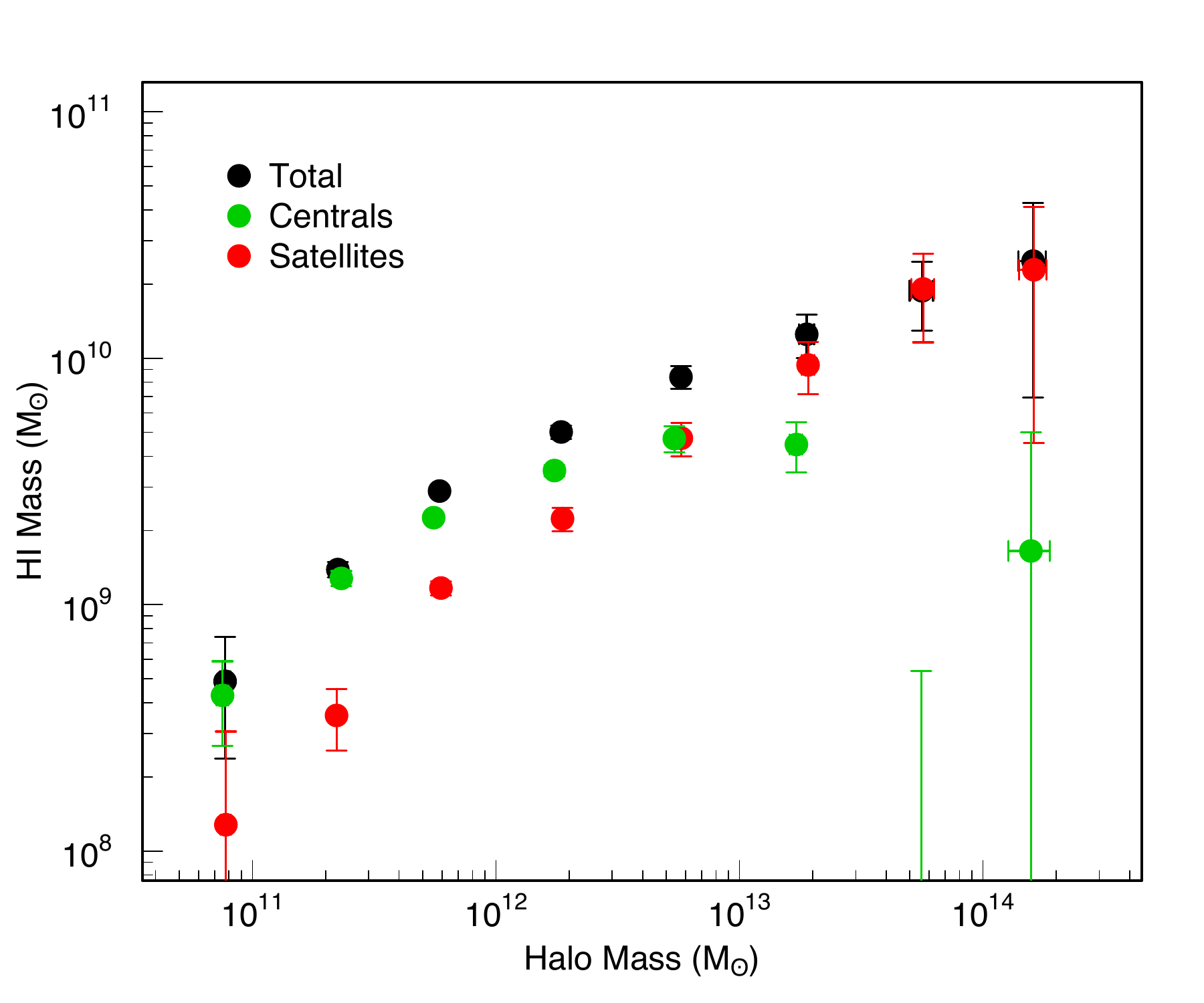}
    \includegraphics[width=0.49\linewidth]{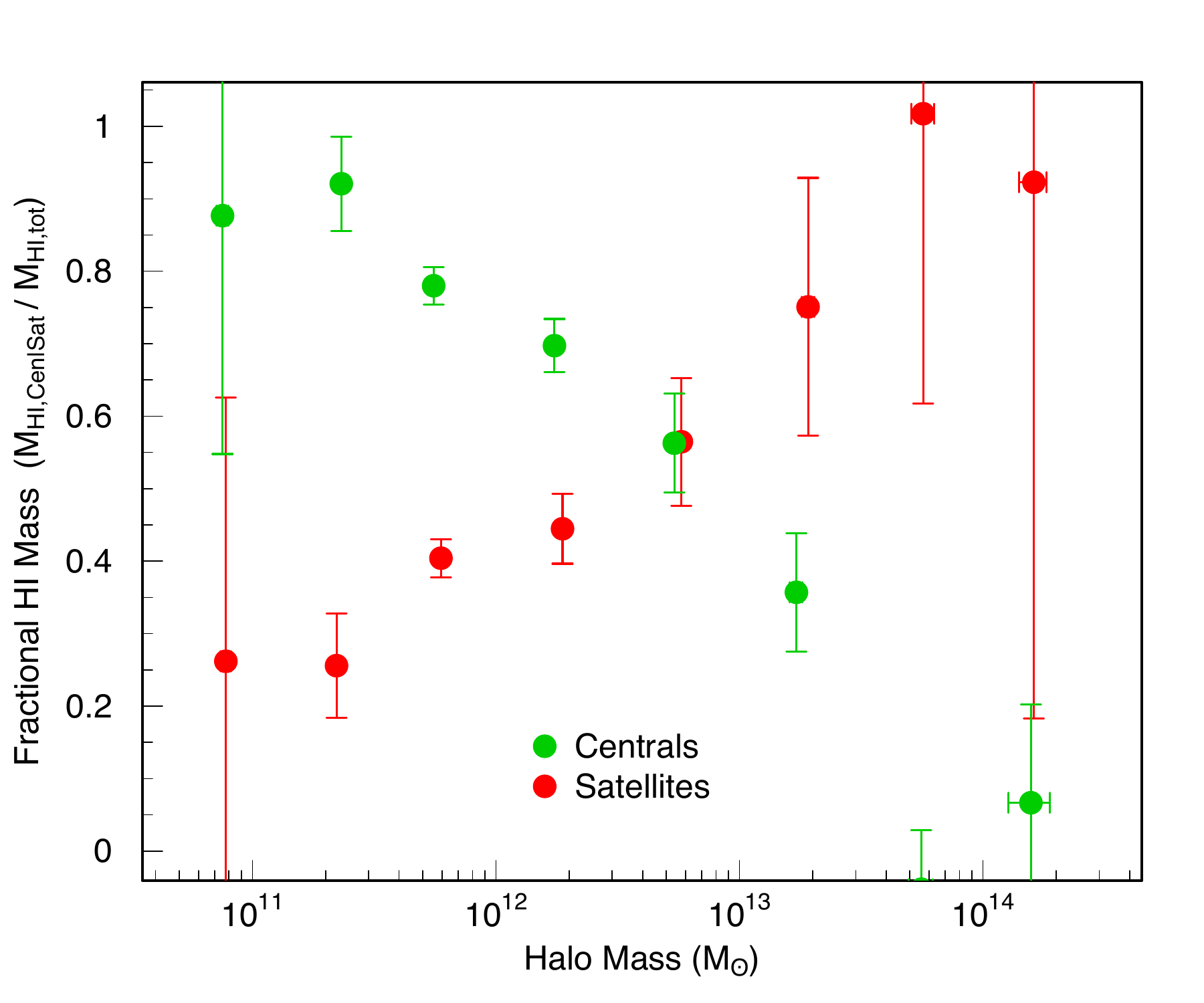}
    \caption{Left: HIHM relation separated in terms of centrals (green), satellites (red) and total (black). Right: Fractional contribution of the centrals and satellites to the total \HI\ mass. The centrals dominate the group \HI\ content up to $M_\text{h} = 10^{13} \text{ M}_\odot$ after which the total satellite \HI\ content starts to dominate.}
    \label{fig:HIHM_Centrals_Satellites}
\end{figure*}

\begin{figure}
    \centering
    \includegraphics[width=\linewidth]{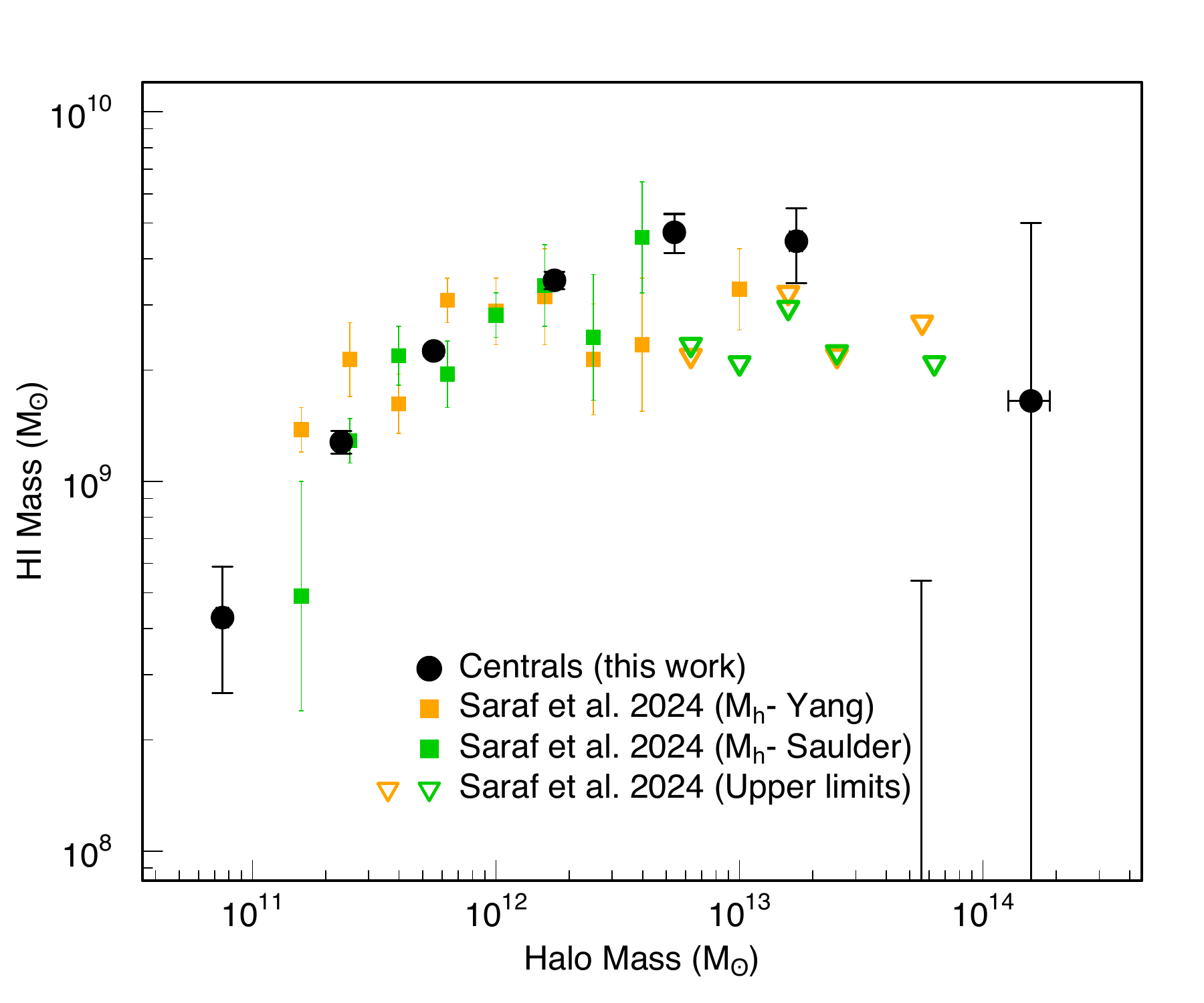}
    \caption{Comparison of our HIHM for centrals (black) with the studies using xGASS sample \citep{Saraf2024} in combination with halo masses from \citet{Yang2007} (orange) and \citet{Saulder2016} (green).}
    \label{fig:Centrals_comparison}
\end{figure}

\textcolor{black}{In Fig. \ref{fig:spectra}, we show the stacked spectra for the AM GAMA and AM GAMA+WAVES samples in 9 halo mass bins. The AM GAMA and AM GAMA+WAVES spectra are consistent for $M_\text{h} < 10^{13} \text{ M}_\odot$, and above this we can see the increased flux due to the inclusion of photometric members in the AM GAMA+WAVES sample. The AM GAMA+WAVES spectra clearly has higher noise compared to AM GAMA sample due to inclusion of photometric interlopers.} 

In Fig. \ref{fig:HIHM_GAMA}, we present stacked \HI\ masses as a function of halo mass for spectroscopic galaxies only from GAMA. Blue open points use velocity dispersion based halo mass estimates, while filled points use AM halo masses. The difference between the AM and velocity dispersion based measurements are due to differences in selection and estimates of the halo mass. For the velocity dispersion based points we select groups with $N_\text{FoF} > 4$ while AM stacks do not have a richness criterion. With the $N_\text{FoF} > 4$ condition, we have 20 groups in G3C above $M_\text{h} > 10^{12} \text{ M}_\odot$ with a total of 190 galaxies. In the case of AM, we have 1082 groups (including 864 isolated galaxies) spanning an AM halo mass range of $10^{10.5}-10^{14.5} \text{ M}_\odot$. 

In Fig. \ref{fig:HIHM_waves}, we present the stacking results comparing the GAMA (blue) and GAMA+WAVES catalog (black). On the left, we show the results for velocity dispersion based halo masses with the criterion $N_\text{FoF}>4$, and on the right we show the stacks with AM based halo masses. With the inclusion of photometric galaxies from WAVES, we recover an average \HI\ content in halos that is greater by a factor of 1.5-3 for high halo masses. However, below $M_\text{h} = 10^{13} \text{ M}_\odot$, there is no significant difference in the group \HI\ content between the GAMA and GAMA+WAVES samples. This is because the satellite abundance  increases with halo mass \citep[e.g.,][]{Carlsten2022, Zhu2023}, and the number of satellites does not vary significantly between the  GAMA and GAMA+WAVES sample for low mass halos. 

This also suggests that a substantial portion of \HI\ in high mass groups and clusters is bound to satellites below the spectroscopic detection threshold. This result highlights the power of including photometric group members - particularly in systems with lower number of spectroscopic members - and cautions against relying solely on spectroscopic samples when assessing total gas content. 

The dip in the HIHM relation around $M_\text{h} = 10^{13.25} \text{ M}_\odot$ in the AM GAMA based stacks  (blue) in Fig. \ref{fig:HIHM_waves} (right) is due to relatively lower satellite count for the groups in that halo mass bin (see Table \ref{tab:halo_counts}). The satellite number count is expected to monotonically increase with halo mass. However, the average $N_\text{FoF}$ as a function of AM halo masses for our GAMA sample does not increase monotonically. This could be due to a combination of uncertainties in AM halo mass estimation and/or spectroscopic incompleteness, or simply a statistical variation. Indeed, the inclusion of the satellite galaxies with the GAMA+WAVES catalog solves this issue as can be seen in black. 

\citet{Rhee2023} observed a dip in the halo mass range $M_\text{h} = 10^{12}-10^{13}\text{ M}_\odot$ using DINGO early science data. Group stacking measurements of \citet{Guo2020} observe a plateau in the same halo mass range, although only when their measurements are restricted to rich systems (over 3 members), which they explain using a three-phase formation scenario of the \HI-rich galaxies. Recently, \citet{Hutchen2025} also observed a shallow valley at $M_\text{h} \sim10^{13}\text{ M}_\odot$ in their study of HIHM relation using RESOLVE and ECO surveys. They also report the presence of a deeper valley ($\sim 0.25 \text{ dex}$) between $M_\text{h} = 10^{11.5}\text{ M}_\odot$  and $M_\text{h} = 10^{12.1}\text{ M}_\odot$, which is not seen in our measurement. A dip in the HIHM relation has been predicted by semi-analytic models like \textsc{Shark} between $M_\text{h} = 10^{12}-10^{13}\text{ M}_\odot$, attributed to the effect of AGN feedback \citep{Chauhan2020, Chauhan2021}. However, we do not find any dip or non-montonic trend in our HIHM relation based on the GAMA+WAVES sample.

We compare our GAMA+WAVES AM HIHM relation with the measurements of \citet{Hutchens2023} in Fig. \ref{fig:HIHM_comparison}. While our measurements are based on stacking, \citet{Hutchens2023} group-integrated \HI\ mass is calculated by summing galactic \HI\ mass estimates from a combination of clean \HI\ detections, strong upper limits, deconfused \HI\ detections, and estimates from photometric gas fractions from the ECO and RESOLVE survey. Their halo masses are based on abundance matching between group halo mass and group integrated r-band luminosity. The HIHM relations derived are consistent with our stacking results.

\subsection{Diffuse or Unaccounted HI}

In Fig. \ref{fig:HIHM_comparison}, we compare  the total group \HI\ measurements of \citet{Guo2020} and \citet{Dev2023} with the galaxy \HI\ measurement in this work. In the group spectral stacking technique, a single spectrum is extracted per group at the group centre by matching the extraction radius with the group radius and spectral velocity range proportional to the group velocity dispersion. This method captures the total \HI\ emission from the group including the \HI\ content of the galaxies plus any diffuse emission. The measurements in \citet{Dev2023} are higher than \citet{Guo2020} primarily because of the $N_\text{FoF}$ selection. \citet{Dev2023} stacks groups with at least two members while \citet{Guo2020} includes systems with one member as well. The HIHM relation shifts up as you increase the $N_\text{FoF}$ and this effect is greater at low halo masses \citep{Guo2020}.

The group stacking measurements are consistent with the galaxy stacks across all the halo masses. This indicates that almost all of the \HI\ mass in a halo can be accounted for by the \HI\ bound to galaxies. Evidence for the presence of dark \HI\ clouds has increased with the advent of more sensitive radio interferometers like MeerKAT and ASKAP. \textcolor{black}{Using MeerKAT, \citet{Josza2022} discovered a massive chain of \HI\ with a total \HI\ mass of $10^{10} \text{ M}_\odot$ which is part of a GAMA galaxy group with a dynamical mass of $10^{13.5} \text{ M}_\odot$. The origin of these \HI\ clouds remains uncertain and is likely related to tidal debris within the group environment. More generally, \citet{O'Beirne2025} demonstrated that many \HI\ “dark clouds” identified in ASKAP surveys are in fact associated with low–surface-brightness galaxies, which are difficult to detect in optical imaging and would typically be missed even by forthcoming deep spectroscopic surveys such as WAVES.} However, our analysis shows that low surface brightness galaxies and dark clouds contribute, on average, insignificantly to the total halo \HI\ mass.

While the group and galaxy stacking measurements are in agreement with each other, it is important to note that there could be systematics in comparison between total group and galaxy \HI\ measurement. The group spectral stacking and galaxy spectral stacking methods are different in terms of their aperture, level of confusion and noise. There are systematics arising from the SMF and HMF used in the abundance matching relation (see Appendix. \ref{sec:appendix_smhm}). Using the \citet{Tinker2008} HMF instead of \citet{Driver2022hmf} HMF, we get systematically higher \HI\ values at lower halo masses ($<10^{12} \text{ M}_\odot$) as shown in Fig. \ref{fig:HIHM_comparison_tinker}. 

There could also be differences due to comparisons between single dish and interferometric observations. Group stacking measurements in \citet{Guo2020} and \citet{Dev2023} have been performed using single-dish observations from ALFALFA whereas the galaxy stacking measurements are using ASKAP interferometric data. \textcolor{black}{\citet{Westmeier2022} compared the flux difference between ASKAP WALLABY public data release 1 and single-dish measurements. The total flux for bright galaxies is accurately recovered while a flux deficit was observed for fainter sources when compared with ALFALFA measurements. Flux deficits of $\sim15\%$ were reported in WALLABY pilot survey phase 2, primarily attributed to insufficient cleaning  \citep{Murugeshan2024}. The impact of insufficient cleaning on source fluxes is partially addressed through the application of the flux correction factor from Rhee et al. (2026, subm.) derived from a comparison of traditional interferometric imaging and image-domain combination. However, we acknowledge that for faint sources there will still likely be an underestimate of source fluxes compared to single dish studies due to insufficient cleaning, as highlighted in the WALLABY studies. As there are only two directly detected HIPASS sources in the field, we are unable to directly estimate this correction factor further for DINGO. However, because the majority of our galaxies are unresolved at the ASKAP resolution, this effect is likely to be minimal in our stacking analysis.} 

In Fig. \ref{fig:HI_fb}, we show the isolated DINGO detections along with our stacking result. We can see that the \HI\ detections can be as much as 1 dex higher than the mean stacked results, indicating that there is a large scatter in the \HI\ masses of individual systems with similar halo/stellar mass. Among our isolated detections, the \HI\ masses can contribute up to 30\% of the cosmological baryon fraction ($f_\text{b,c}$) whereas the average \HI\ contribution is $\sim5\%$ near the turnover ($M_\text{h} \sim 10^{11} \text{ M}_\odot$) in the HIHM relation. 

We fit the stacked results with a double power-law model of the form,
\begin{equation}
    \mathrm{log_{10}}(M) = N - \mathrm{log_{10}} \left[ \left( \frac{10^{\mathcal{M}_h}}{10^{\mathcal{M}_o}}\right)^{-\alpha} + \left( \frac{10^{\mathcal{M}_h}}{10^{\mathcal{M}_o}}\right)^{-\beta}\right] ,
\end{equation}
where $\mathcal{M}_h = \mathrm{log_{10}}(M_h/\text{M}_\odot)$, $M$ is the baryonic component mass, $\alpha \text{ and } \beta$ are the two exponents of the power-law, $\mathcal{M}_o$ is the turnover halo mass which separates the two different power-law regimes, N is the normalization parameter. The best-fit parameters are provided in Table. \ref{tab:HIHM_params}. The turnover mass reported in \citet{Hutchens2023} ($10^{11.4-11.5} \text{ M}_\odot$) is higher but within our uncertainties. This could be due to insufficient sampling below $M_\text{h} = 10^{11} \text{ M}_\odot$ in both the datasets. We also performed a single power-law fit. In Fig. \ref{fig:HI_fb}, we show both the fits; however the data favour a double power-law.

\subsection{Centrals and Satellites}

In Fig. \ref{fig:HIHM_Centrals_Satellites} (left panel), we show the H I–halo mass (HIHM) relation from the GAMA+WAVES sample, separating the contributions of centrals and satellites. Centrals are the BCGs or isolated centrals defined in the G3C, while satellites include all galaxies assigned to a given group excluding the BCG. Stacking was performed separately for centrals (green circles), satellites (red circles), and the total sample (black circles). Due to stacking systematics, the sum of the central and satellite \HI\ masses does not necessarily match the total exactly, but they agree within the uncertainties. The suppressed \HI\ content of centrals in high-mass halos is consistent with the known trends in star-forming activity. For example, \citet{Kauffmann2003} showed that star formation becomes increasingly rare in high stellar mass systems, and \citet{Fraser-McKelvie2014} found that majority of the cluster centrals in the local universe ($z \le 0.1$) exhibit very little or no ongoing star formation. These results suggest that \HI\ reservoirs in massive centrals are depleted or inefficiently replenished, naturally leading to a declining central \HI\ fraction with increasing halo mass.

In the right panel of Fig. \ref{fig:HIHM_Centrals_Satellites}, we show the fractional contributions of centrals and satellites to the total group H I mass. The total group \HI\ content is dominated by centrals at low halo mass, and by satellites at high halo masses. Centrals dominate up to $M_\text{h} \sim 6\times10^{12} \text{ M}_\odot$ beyond which the cumulative satellite H I content becomes larger. This trend follows the expected increase in satellite abundance with halo mass, and agrees well with the halo-model predictions of \citet{Villaescusa-Navarro2018}, who find that centrals dominate the halo \HI\ fraction in low-mass halos, with satellites becoming dominant above $M_\text{h}=7\times10^{12} \text{ M}_\odot$.

We also compare our HIHM relation for centrals with \citet{Saraf2024}. They used the extended GALEX Arecibo SDSS Survey \citep{Cantinella2018} to study the median HIHM across $11.1 < \log_{10}\left(M_\text{h}/\text{M}_\odot \right) < 14.1$. In Fig. \ref{fig:Centrals_comparison}, we show the comparison of our central HIHM relation to the results of \citet{Saraf2024} using halo masses from \citet{Yang2007} (orange) and \citet{Saulder2016} (green). The results are consistent across all the halo masses with detections. The flatness of the HIHM relation for centrals was attributed to be arising from two different population of centrals at opposite ends of the halo mass distribution. At low halo masses, star-forming disc galaxies that are \HI-rich are more prominent and drive up the \HI\ masses, whereas, the passive, \HI\ -poor, high stellar concentration systems dominating the centrals at high halo masses drive the relation down. 

The total stellar mass contribution of groups shifts from central-dominated to satellite-dominated above $\sim 10^{13} \text{ M}_\odot$ \citep{Shuntov2022, Zacharegkas2025}. So, the halo mass scale where the dominant \HI\ contribution shifts from central to satellite galaxies is slightly below the turnover for the stellar mass. 


\begin{table}
\centering
\renewcommand{\arraystretch}{1.5}
\begin{tabular}{lccc}
\hline
 $\alpha$  & $\beta$ & $\mathcal{M_\text{o}}$ & N \\
\hline
 $0.38 \pm 0.10$ & $1.51 \pm 0.38$ & $11.18 \pm 0.28$ & $9.32 \pm 0.20$\\
\hline
\end{tabular}
\caption{Best-fit double power-law parameters (Eq. 4) for the HIHM relation for the GAMA+WAVES AM sample.}
\label{tab:HIHM_params}
\end{table}

\section{Conclusion}

In this work, we study the \HI\ content in halos as a function of halo mass using the DINGO 100h pilot phase data combined with optical data from the GAMA spectroscopic survey and the WAVES photometric catalog. The DINGO 100h data cube has a restoring beam size of $\sim 30'' \times 30''$ with a mean RMS sensitivity of $\sim 0.67 \text{ mJy beam}^{-1}$.  We have 408 \HI\ detections with optical counterparts ranging from $\sim~10^7 \text{ M}_\odot$ to over $10^{10} \text{ M}_\odot$ in \HI\ mass. 

We use the spectral-stacking to measure the \HI\ content in galaxies within the halos. Spectral stacking is performed in two ways - first by stacking \HI\ spectra of spectroscopically selected galaxies from GAMA and second by stacking the \HI\ spectra of galaxies from the GAMA+WAVES catalog at the redshift of the GAMA central galaxy for each group. \textcolor{black}{The inclusion of photometric WAVES galaxies enables a more complete accounting of satellite \HI\ emission that would otherwise be missed due to spectroscopic incompleteness. Although these galaxies lack spectroscopic redshifts, the stacking is performed at the spectroscopic redshift of the group central, such that true satellite emission remains clustered in velocity space within the halo velocity dispersion and therefore adds coherently to the stacked signal. Projected interlopers contribute primarily as additional noise, leading to a reduction in the signal-to-noise of the resulting stack. However, the net effect is an improved estimate of satellite \HI\ emission and a more complete census of the total group \HI\ content.}

We obtain higher average group \HI\ content in halos above $M_\text{h}\sim10^{13} \text{ M}_\odot$ when fainter photometric galaxies are used to supplement the spectroscopic galaxy sample by a factor of 1.5-3. This indicates the presence of significant \HI\ content in satellite galaxies below the completeness limit of G23 (r < 19.42 mag). For lower mass halos, we do not find such a difference. This is expected, since the number of satellites in low mass halos is small \citep{Carlsten2022, Zhu2023}. 

The galaxy spectral stacking results are consistent with the group spectral stacking results of \citet{Guo2020} in the group and cluster regime. \textcolor{black}{This suggests that any additional contribution to the total halo \HI\ mass from dark clouds or low-surface-brightness galaxies that fall below the optical detection limit—and are therefore not included in our stacking sample—is likely to be small.}

We also divide our sample into centrals and satellites to study the HIHM relation separately for the two samples. The fraction of \HI\ in central galaxies decreases with halo mass while the fraction of \HI\ in satellite galaxies increases. At $M_\text{h}\sim 6 \times 10^{12} \text{ M}_\odot$, the fractional contribution by centrals and satellites are almost same.

This work presents a preliminary study of the HIHM relation with the DINGO 100h pilot data, and the full DINGO survey will deliver nearly 800 hours of observations over two comparable sky tiles at the same redshift range, substantially improving sensitivity and sample statistics in the future. Our novel stacking technique that associates photometric members with a spectroscopically identified group central galaxy to detect possible \HI\ emission from group members will be very useful in the future with the influx of data from large optical imaging and spectroscopic surveys using LSST \citep{Ivezic2019}, Euclid \citep{EUCLID}, Roman \citep{ROMAN}, DESI \citep{DESI} and 4MOST \citep{deJong2019} when combined with deeper \HI\ surveys from ASKAP \citep{Hotan2021}, MeerKAT \citep{MeerKAT}, FAST (Five-hundred-meter Aperture Spherical radio Telescope; \citealt{FAST}) and eventually SKA (Square Kilometre Array; \citealt{Staveley-Smith2015}).

\section*{Acknowledgements}

We thank the referee for their useful comments that helped to improve the manuscript. AD acknowledges the support from University Postgraduate Award and Scholarship for International Research Fees from the University of Western Australia. SPD acknowledges financial support via an Australian Research Council (ARC) Australian Laureate Fellowship award (FL220100191). ET acknowledges funding from the HTM (grant TK202), ETAg (grant PRG3034) and the EU Horizon Europe (EXCOSM, grant No. 101159513).

This work was supported by resources provided by the Pawsey Supercomputing Research Centre’s Setonix Supercomputer (https://doi.org/10.48569/18sb-8s43), with funding from the Australian Government and the Government of Western Australia.

This scientific work uses data obtained from Inyarrimanha Ilgari Bundara, the CSIRO Murchison Radio-astronomy Observatory. We acknowledge the Wajarri Yamaji People as the Traditional Owners and native title holders of the Observatory site. CSIRO’s ASKAP radio telescope is part of the Australia Telescope National Facility (https://ror.org/05qajvd42). Operation of ASKAP is funded by the Australian Government with support from the National Collaborative Research Infrastructure Strategy. ASKAP uses the resources of the Pawsey Supercomputing Research Centre. Establishment of ASKAP, Inyarrimanha Ilgari Bundara, the CSIRO Murchison Radio-astronomy Observatory and the Pawsey Supercomputing Research Centre are initiatives of the
Australian Government, with support from the Government of Western Australia and the Science
and Industry Endowment Fund. 

GAMA is a joint European-Australasian project based around a spectroscopic campaign using the Anglo-Australian Telescope. The GAMA input catalogue is based on data taken from the Sloan Digital Sky Survey and the UKIRT Infrared Deep Sky Survey. Complementary imaging of the GAMA regions was obtained by a number of independent survey programmes including GALEX MIS, VST KiDS, VISTA VIKING, WISE, Herschel-ATLAS, GMRT and ASKAP providing UV to radio coverage. GAMA is funded by the STFC (UK), the ARC (Australia), the AAO, and the participating institutions. The GAMA website is https://www.gama-survey.org/ . Based on observations made with ESO Telescopes at the La Silla Paranal Observatory under programme ID 179.A-2004. Based on observations made with ESO Telescopes at the La Silla Paranal Observatory under programme ID 177.A-3016.

WAVES is a joint European-Australian project based around a spectroscopic campaign using the 4-metre Multi-Object Spectroscopic Telescope. The WAVES photometric catalog is based on data taken from the European Southern Observatory’s VST and VISTA telescopes (ID 179.A-2004 and ID 177.A-3016). Complementary imaging of the WAVES regions is being obtained by a number of independent survey programmes including GALEX MIS, VST, WISE, Herschel-ATLAS, and ASKAP providing UV to radio coverage. WAVES is funded by the ARC (Australia) and the participating institutions. The WAVES website is https://wavesurvey.org.\\

\section*{Data Availability}
The data underlying this article will be shared upon reasonable request to the corresponding author, AD.

\bibliographystyle{mnras}
\bibliography{HIGroups_main}

\appendix

\section{Galaxy Stellar Mass Function for centrals}
\label{sec:appendix_gsmf}

\begin{table*}
\centering
\renewcommand{\arraystretch}{1.5}
\begin{tabular}{lcccccc}
\hline
Sample & $\log_{10}(M_*)$ & $\log_{10}(\phi_1)$ & $\log_{10}(\phi_2)$ & $\alpha_1$ & $\alpha_2$ \\
       & [M$_\odot$]      & [Mpc$^{-3}$]        & [Mpc$^{-3}$]        &            &            \\
\hline
All        & $10.75 \pm 0.02$ & $-2.44 \pm 0.02$ & $-3.20 \pm 0.06$ & $-0.47 \pm 0.07$ & $-1.53 \pm 0.03$ \\
Centrals   & $10.76 \pm 0.02$ & $-2.57 \pm 0.02$ & $-3.40 \pm 0.07$ & $-0.39 \pm 0.07$ & $-1.53 \pm 0.03$ \\
Satellites & $10.64 \pm 0.04$ & $-2.93 \pm 0.03$ & $-3.67 \pm 0.17$ & $-0.56 \pm 0.16$ & $-1.57 \pm 0.07$ \\
\hline
\end{tabular}
\caption{Best-fit double Schechter parameters and $1\sigma$ uncertainties for each sample. $\phi_1$, $\phi_2$ and $\alpha_1$, $\alpha_2$ describe the vertical normalization and slope parameters, respectively, for the two components, and $M_*$ is the characteristic stellar mass that acts as the horizontal normalization.}
\label{tab:SMF_params}
\end{table*}

\begin{figure}
    \centering
    \includegraphics[width=\linewidth]{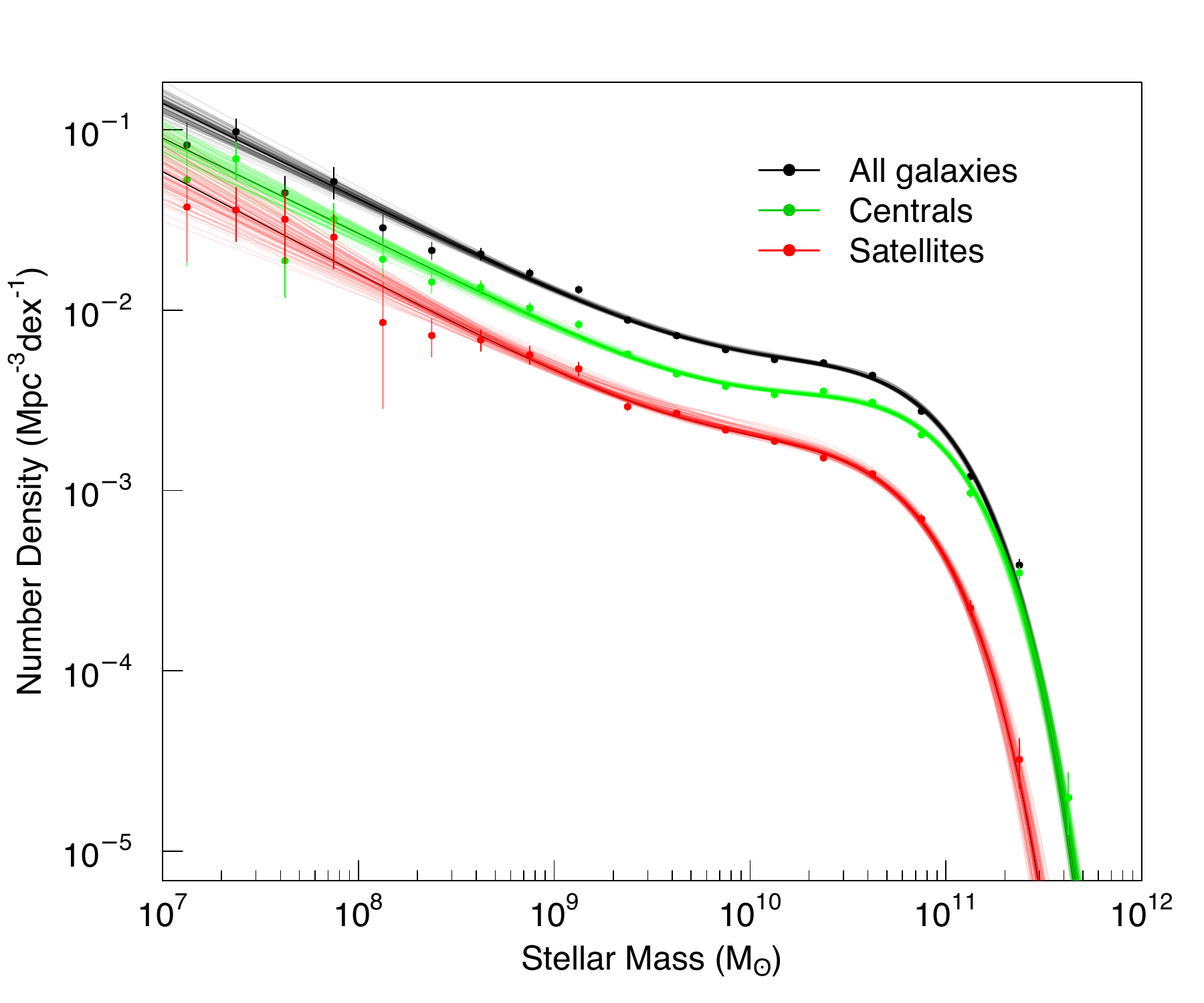}
    \caption{Galaxy stellar mass function for the GAMA galaxies with z<0.1. The full sample (black) is split into two - centrals (green) which includes BCG and isolated galaxies, and satellites (red).}
    \label{fig:GSMF}
\end{figure}

We follow the methodology adopted in \citet{Driver2022dr4} to construct the SMF for centrals using GAMA galaxies with $z<0.1$. The method uses the Modified Maximum Likelihood estimation, accessed through \textsc{DFTOOLS} \citep{Obreschkow2018}, to fit a double Schechter function to fit our volume-corrected galaxy number density as a function of stellar mass. We use the same galaxy sample used in \citet{Driver2022dr4} to construct the SMF, which includes galaxies from G09, G12, G15 and G23. The sample has a redshift completeness of over 95\% for an r-band magnitude limit of 19.65 mag. We divide the galaxy sample into centrals and satellites. The centrals include all the BCGs from the G3C and isolated galaxies that are not part of any GAMA group. The satellites include all galaxies in groups as defined by the G3C excluding the centrals. We fit a double-schecter model to the SMF for all the three samples - total, centrals and satellites. The double-schecter form in logarithmic mass intervals is given by:
\begin{equation}
\begin{aligned}
\phi\left(\log_{10}\left(\frac{M}{M_\odot}\right)\right) &\equiv \frac{dn}{d\left(\log_{10}\left(M/M_\odot\right)\right)} \\
&= \log(10) \, \exp\left[-\frac{M}{M^*}\right] \\
&\quad \times \left[ \phi_1 \left(\frac{M}{M^*}\right)^{\alpha_1 + 1} + \phi_2 \left(\frac{M}{M^*}\right)^{\alpha_2 + 1} \right]
\end{aligned}
\end{equation}
where $\phi_1$, $\phi_2$ and $\alpha_1$, $\alpha_2$ describe the vertical normalization and slope parameters, respectively, for the two components, and $M_*$ is the characteristic stellar mass that acts as the horizontal normalization. In Fig. \ref{fig:GSMF}, we show SMF for centrals (green), satellites (red) and the full sample (black). Our full sample results are consistent, as expected, with the \citet{Driver2022dr4} GSMF. The best-fit parameters for all the three fits are given in Table \ref{tab:SMF_params}. For our abundance matching, we use the SMF constructed for the centrals.

\begin{figure}
    \centering
    \includegraphics[width=\linewidth]{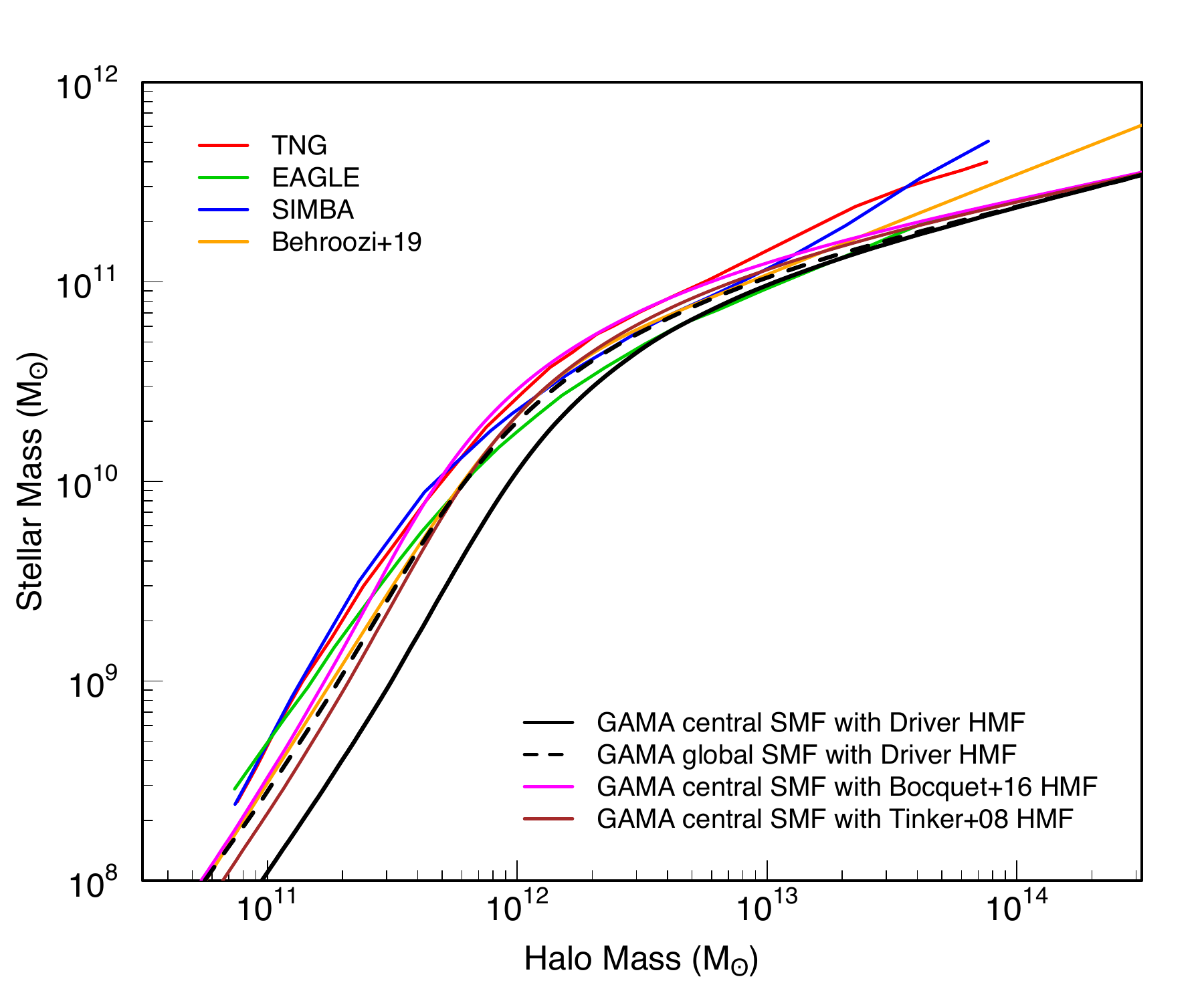}
    \caption{Stellar to halo mass relation constructed from abundance matching using our central SMF and \citet{Driver2022hmf} HMF with Planck-constraint (black solid). We also show SMHM relation using our global SMF and \citet{Driver2022hmf} HMF with Planck-constraint (black dashed). Variations using different theoretical HMFs are shown using \citet{Tinker2008} (brown) and \citet{Bocquet2016} HMF in combination with GAMA central SMF. We also show SMHM relation for centrals from TNG (red), EAGLE (green), SIMBA (blue) and \citet{Behroozi2019} (orange) models. }
    \label{fig:SMHM}
\end{figure}

\section{Sensitivity to SMHM relation}
\label{sec:appendix_smhm}
\begin{figure}
    \centering
    \includegraphics[width=\linewidth]{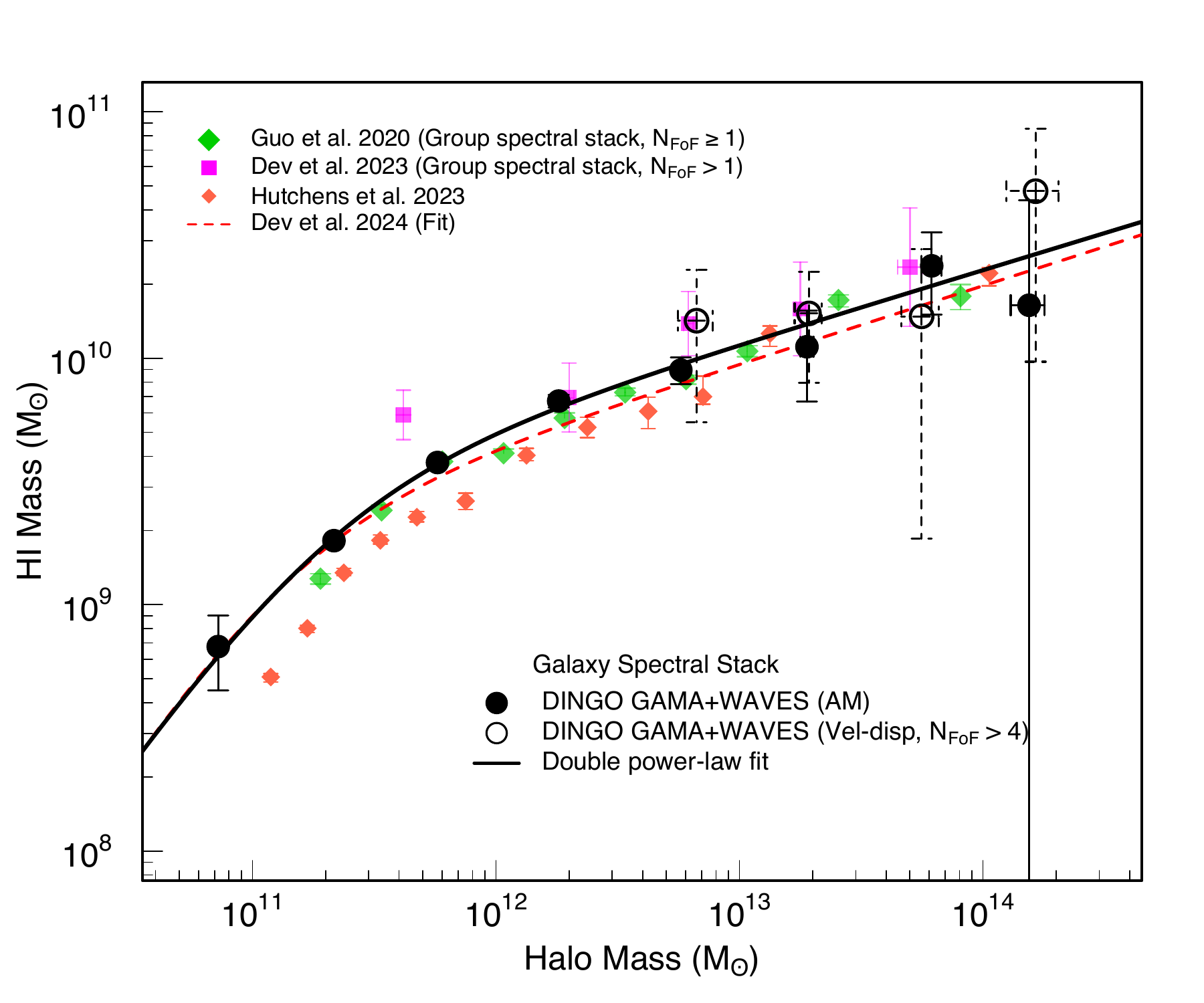}
    \caption{Same as Fig. \ref{fig:HIHM_comparison}, but AM performed with \citet{Tinker2008} HMF.}
    \label{fig:HIHM_comparison_tinker}
\end{figure}

Halo masses for our analysis are primarily obtained using abundance matching, which is based on the SMF and HMF. In Fig. \ref{fig:SMHM}, we show the SMHM relation used in this work which is obtained using GAMA SMF for centrals and the \citet{Driver2022hmf} HMF with the Planck $\Omega_\text{M}$ constraint, that imposes the integral under the HMF to be $\Omega_\text{M}=0.31$. We also show SMHM relations constructed using the global GAMA SMF and \citet{Driver2022hmf} HMF (black). The two relations are consistent at the high mass end, because the global and central SMF are consistent at the high stellar mass end. But at low stellar masses, the satellite contribution to the global SMF is not negligible. Hence, a central at fixed stellar mass will be assigned a lower halo mass when using a global SMF instead of central SMF. We show other SMHM relation using GAMA SMF for centrals, and HMFs from \citet{Tinker2008} and \citet{Bocquet2016} corrected to the same cosmology. Also shown are SMHM relation for centrals from EAGLE, TNG and SIMBA simulations \citep{Wright2024}. There is an offset between the SMHM relations between the \citet{Driver2022hmf} and theoretical HMFs. This could be the result of the Planck-constraint imposed in \citet{Driver2022hmf} or an underestimate of the Eddington-bias correction.  Hence, the abundance matched SMHM relation is sensitive to both the SMF and HMF. This will in-turn also affect the HIHM relation and is one of the main systematic uncertainties. In Fig. \ref{fig:HIHM_comparison_tinker}, we show HIHM relation, similar to Fig. \ref{fig:HIHM_comparison}, but uses abundance matched relation constructed from GAMA central SMF and \citet{Tinker2008} HMF. This can be resolved with better empirically constrained HMF pushing down to lower halo masses, one of the science goals of the 4MOST WAVES survey. 



\begin{table*}
\centering
\renewcommand{\arraystretch}{1.5}
\begin{tabular}{ccccccccc}
\hline
$\log_{10}(M_\text{h}/\text{M}_\odot)$ & \multicolumn{2}{c}{AM GAMA} & \multicolumn{2}{c}{AM GAMA+WAVES} & \multicolumn{2}{c}{Vel-disp GAMA} & \multicolumn{2}{c}{Vel-disp GAMA+WAVES} \\
                 & $N_{\rm gal}$ & $N_{\rm group}$ & $N_{\rm gal}$ & $N_{\rm group}$ & $N_{\rm gal}$ & $N_{\rm group}$ & $N_{\rm gal}$ & $N_{\rm group}$ \\
\hline
10.5--11   & 15  & 15  & 102   & 15  & -- & -- & --  & -- \\
11--11.5   & 192 & 183 & 846  & 183 & -- & -- & --  & -- \\
11.5--12   & 605 & 538 & 2429 & 538 & -- & -- & --  & -- \\
12--12.5   & 324 & 225 & 1928 & 225 & -- & -- & --  & -- \\
12.5--13   & 214 & 79  & 1498 & 79  & 16 & 2  & 79 & 2  \\
13--13.5   & 61  & 25  & 872  & 25  & 81 & 11  & 287 & 11  \\
13.5--14   & 83  & 12  & 670  & 12  & 75 & 6  & 293 & 6  \\
14--14.5   & 18  & 5   & 461  & 5  & 18 & 1  & 71  & 1  \\
\textbf{Total} 
& \textbf{1512} & \textbf{1082} 
& \textbf{8806} & \textbf{1082} 
& \textbf{190}  & \textbf{20} 
& \textbf{730}  & \textbf{20} \\
\hline
\end{tabular}
\caption{Galaxy and group counts for different halo mass bins and samples used in the stacking analysis. The AM halo mass based sample includes groups with $N_\text{FoF}\ge1$, while the Vel-disp halo mass based sample includes groups with $N_\text{FoF}\ge5$. 
}
\label{tab:halo_counts}
\end{table*}

\label{lastpage}
\end{document}